\documentclass[prl,aps,amsfonts,preprint,floats,tightenlines,floatfix]{revtex4}
\usepackage{ifthen}                   % needed for Feyman-Slash macro
\usepackage{exscale}                  % correct scaling of math-symbols
\usepackage[intlimits]{amsmath}       % improved mathematical typesetting
\usepackage{amsfonts}
\usepackage{amssymb,amscd}
\usepackage[dvips]{epsfig}
\usepackage{array}
\usepackage{wrapfig}

\newcommand{\beqa}{\begin{eqnarray}}
\newcommand{\eeqa}{\end{eqnarray}}
\newcommand{\beq}{\begin{equation}}
\newcommand{\eeq}{\end{equation}}
\newcommand{\pslash}{p\!\!\!/\,}
\newcommand{\qslash}{q\!\!\!/\,}

\begin{document}

\date{\today}
\title{
\hspace*{\fill}{\small\sf UNITU--THEP--26/2002}\\
\hspace*{\fill}{\small\sf http://xxx.lanl.gov/abs/hep-ph/0301094}\\ ~\\ ~\\
Non-perturbative Propagators, Running Coupling and\\ Dynamical Quark Mass
of Landau gauge QCD}
\author{
C.~S.~Fischer\footnote{Supported by the European Graduate School
Basel-Tuebingen\\ E-Mail: chfi@tphys.physik.uni-tuebingen.de\\
} and
R.~Alkofer\footnote{E-Mail: reinhard.alkofer@uni-tuebingen.de}}
\affiliation{Institute for Theoretical Physics, University of T\"ubingen \\
          Auf der Morgenstelle 14, D-72076 T\"ubingen, Germany}
~\\
~\\

\begin{abstract}
The coupled system of renormalized  Dyson--Schwinger equations for the quark,
gluon and ghost propagators of  Landau gauge QCD is solved within truncation
schemes. These employ bare as well as non-perturbative ans{\"a}tze for the vertices
such that the running coupling as well as the quark mass function are
independent of the renormalization point. 
The one-loop anomalous dimensions of all propagators are reproduced. Dynamical
chiral symmetry breaking is found, the dynamically generated quark mass agrees
well with phenomenological values and corresponding results from lattice
calculations. The effects of unquenching the system are small. In
particular the infrared behavior of the ghost and gluon dressing functions
found in previous studies is almost unchanged as long as the number of light
flavors is smaller than four.
~
\end{abstract}
\maketitle

\section{Introduction}

Based on many observations in hadron physics spontaneous breaking of chiral
symmetry and the dynamical generation of quark masses are expected to occur in
Quantum Chromo Dynamics (QCD). The precise origin of this non-perturbative
phenomenon as well as its relation to quark confinement are still little
understood. Further studies of these issues have to build on reliable
non-perturbative methods. And, as confinement is expected to be correlated
with  infrared singularities, continuum-based methods will be required in
addition to  Monte Carlo lattice calculations. To this end we note that the
Schwinger--Dyson equations (DSEs) of QCD can address directly the infrared
region of momentum.

The DSEs for the propagators of QCD form a coupled system of equations. In
Landau gauge these have been investigated in two lines of research. On the one
hand, the DSEs of pure Yang--Mills theory have been explored with the aim to
reveal the infrared behavior of the ghost and gluon propagator and their
relation to gluon confinement (see e.g.\ the review~\cite{Alkofer:2000wg} and
references therein). On the other hand, the quark DSEs have been studied
extensively on the purpose of model building. These models have been used in
the frameworks of Bethe--Salpeter equations and finite temperature field theory
to describe hadronic properties and reactions in a semi-phenomenological way
(see e.g.\ the review \cite{Roberts:2000aa}
and  references therein).

In this article numerical results for the coupled set of quark, gluon {\bf and}
ghost propagators, including the backreaction of the quarks on the ghosts and
gluons, are presented for the first time. In the quark DSE we will study the
mechanism of dynamical chiral symmetry breaking by which physical quark masses
are generated even though the bare quark masses in the Lagrangian are zero.
This is a genuine non-perturbative effect as it is well known that for
vanishing bare masses the renormalized masses remain zero at each order in
perturbation theory. In addition to the phenomenon of mass generation we are
interested in quark confinement. Single quark states have non-vanishing color
charge and are therefore not contained in the physical part of the state space
of QCD. This physical subspace supports a positive (semi-)definite metric 
whereas the remaining state space of QCD contains negative norm states as well.
Consequently, negative norm contributions to the quark propagator would provide
evidence for quark confinement.

This paper is organized as follows: In the next section suitable {\it
ans\"atze} for the quark-gluon vertex will be constructed such that the DSE for
the quark propagator guarantees the realization of two important consistency
conditions. These are (i) the  independence of the dynamically generated quark
mass function from the  renormalization point and (ii) the correct asymptotic
behavior at large momenta such that the anomalous dimensions of dressing and
mass functions are correct in one-loop order. Fortunately, the corresponding
DSEs for the fermions of QED are well studied (a short overview is given {\it
e.g.\/} in \cite{Pennington:1998cj}). We will dwell on these results and
construct non-Abelian generalizations of Abelian vertices, which have the
desired properties.

In the following section we present solutions for the quenched system of quark,
ghost and gluon DSEs, {\it i.e.}~we neglect the quark-loop in the gluon
equation. The Yang--Mills sector is hereby treated in  a truncation scheme for
the ghost and gluon equations that already has been employed in ref.\
\cite{Fischer:2002hn}. This scheme improves on older ones
\cite{vonSmekal:1997is,Atkinson:1998tu}, provides an explicit numerical
solution for the infrared analysis given in refs.\
\cite{Lerche:2002ep,Zwanziger:2001kw}, and its results are in almost
quantitative agreement with corresponding results of lattice calculations for
the gluon and ghost propagators 
\cite{Suman:1996zg,Bonnet:2000kw,Cucchieri:1998fy,Langfeld:2001cz}. The main
purpose of calculating the quark propagator in quenched approximation is to
allow for a comparison with corresponding recent lattice results
\cite{Bonnet:2002ih,Bowman:2002kn}. We find very good agreement of our results 
for the quark renormalization function and the momentum-dependent dynamical 
quark mass with the lattice results if the quark-gluon vertex functions 
constructed in Sect.~2 %\ref{quark-dse-sec} 
are employed. We then proceed to the unquenched
case and incorporate the quark-loop into our truncation scheme for the ghost
and gluon DSE. We present solutions for the full coupled system of DSEs for the
quark, ghost and gluon propagators. Compared to the quenched case we will find
only moderate differences for the number of light flavors $N_f \le 3$. 
Whereas we are able to demonstrate positivity violation in the gluon propagator
(and thus gluon confinement) we have not been able to draw definite conclusions
on the (non-)positivity of the quark propagator. In the last section we
summarize our results and present some conclusions.

\section{The quark Dyson-Schwinger equation }
\label{quark-dse-sec}

\begin{center}
\begin{figure}
  \centerline{ \epsfig{file=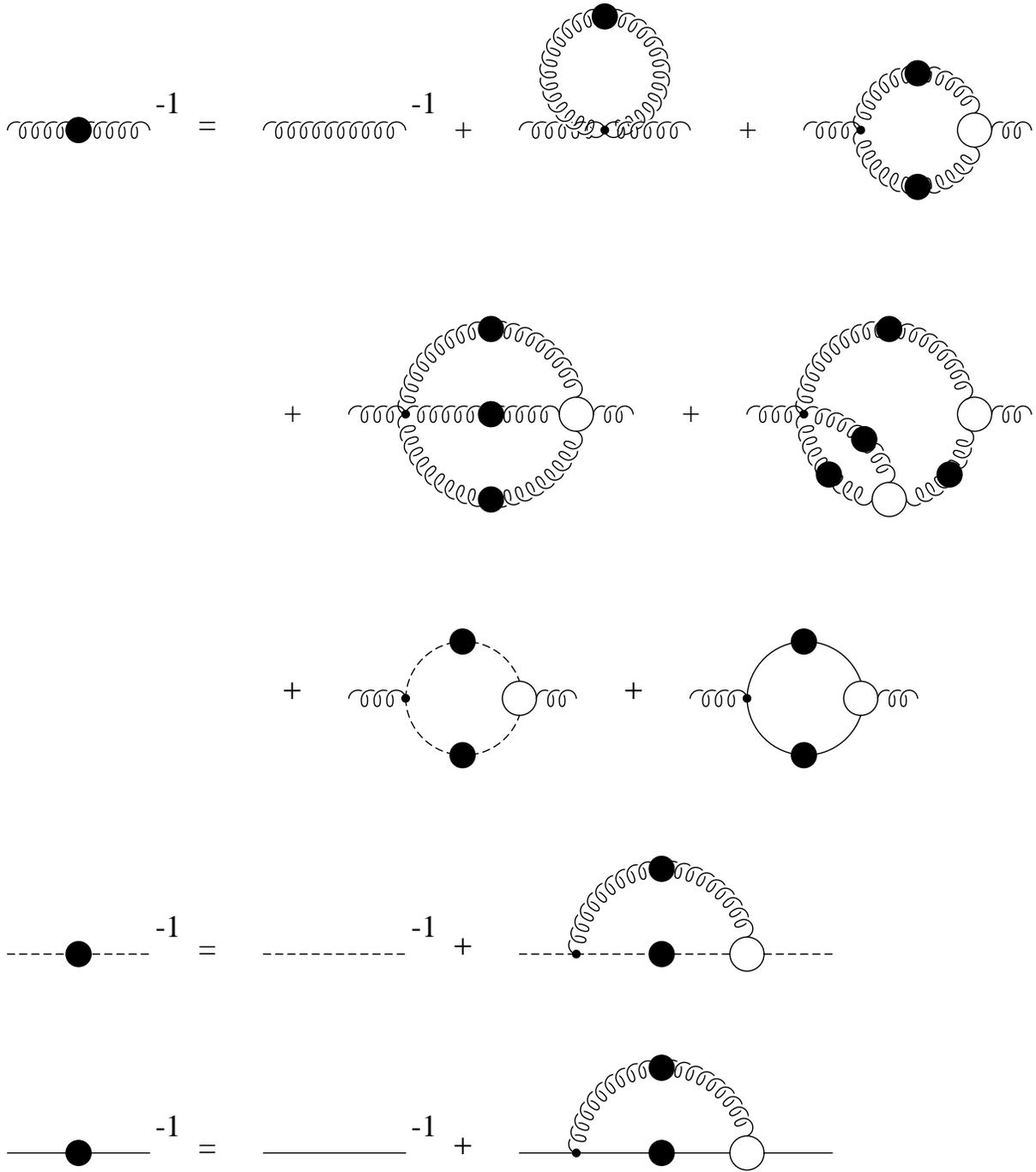,width=0.98\linewidth} }
  \vskip 5mm
  \caption{Diagrammatic representation of the 
           Dyson--Schwinger equations for the gluon, ghost and quark
	   propagators. The wiggly, dashed and solid
           lines represent the propagation of gluons, ghosts and quarks,
           respectively. A filled blob represents a full propagator and
           a circle indicates a one-particle irreducible vertex. }
  \label{GluonGhostQuark}
\end{figure}
\end{center}

The coupled DSEs for the gluon, ghost and quark propagators in Landau gauge QCD
are shown in in fig.\ \ref{GluonGhostQuark}. As stated in the introduction we
will employ first the quenched approximation, {\it i.e.\/} we will neglect the
quark-loop in the gluon equation. This will allow us to assess the quality of our quark DSE
solution by comparing to lattice calculations of the quark propagator
\cite{Bonnet:2002ih} performed so far only in quenched approximation.
As we furthermore employ the gluon propagator obtained in ref.\ 
\cite{Fischer:2002hn} the unknown element to be determined in the quark DSE is
the dressed quark-gluon vertex.

Following the conventions and notations of ref.\ \cite{Alkofer:2000wg} the
renormalized quark DSE with appropriate quark wave function  and quark-gluon
vertex renormalization constants, $Z_2$ and $Z_{1F}$, respectively, reads
\begin{equation}
S^{-1}(p) = Z_2 \, S^{-1}_0(p) + 
{g^2}\, Z_{1F}\, C_F\,
\int \frac{d^4q}{16\pi^4}\, \gamma_{\mu}\, S(q) \,\Gamma_\nu(q,q-p) \,
D_{\mu \nu}(q-p) \, .
\label{quark1} 
\end{equation} 
The factor $C_F=(N_c^2-1)/2N_c$ in front of the integral stems from the color
trace of the loop. The symbol $\Gamma_\nu(q,q-p)$ denotes the full quark-gluon
vertex. 
Suppressing color indices the quark and gluon propagators in Landau gauge 
are given by
\beqa
S(p) &=& \frac{1}{-i \pslash A(p^2) + B(p^2)} \:\:\:
= \:\:\: A^{-1}(p^2)\frac{i \pslash + M(p^2)}{p^2 + M^2(p^2)} \,, \\
S_0(p) &=& \frac{1}{-i \pslash + m_0}
\,, \\
D_{\mu \nu}(p) &=& \left(\delta_{\mu \nu} - \frac{p_\mu p_\nu}{p^2} \right) 
\frac{Z(p^2)}{p^2} \,,
\eeqa
where the quark mass function $M$ is defined as $M(p^2)=B(p^2)/A(p^2)$.

The inverse of the vector self-energy, $1/A(q^2)$, is often denoted as the
'quark wave function renormalization'. The bare quark propagator $S_0(p^2)$
contains the unrenormalized quark mass $m_0(\Lambda^2)$ which depends on the
cutoff $\Lambda$ of the theory. The bare mass is related to the renormalized
mass $m_R(\mu^2)$ via the renormalization constant $Z_m$: 
\begin{equation}
m_0(\Lambda^2)=Z_m(\mu^2,\Lambda^2) \: m_R(\mu^2) \,. 
\label{mass-ren} 
\end{equation}
Here $\mu^2$ is the (squared) renormalization scale. Note that the propagator
functions $A$, $B$ and $Z$ depend on this renormalization scale. Therefore 
in the following we will use the notation $A(p^2,\mu^2)$, $B(p^2,\mu^2)$ and
$Z(p^2,\mu^2)$.

The renormalized and the bare vector self-energy, $A$ and $A_0$, are related 
by
\begin{equation}
A_0^{-1}(p^2,\Lambda^2) = Z_2(\mu^2,\Lambda^2) \: A^{-1}(p^2,\mu^2).
\label{Aren}
\end{equation}
In Landau gauge the loop corrections to the vector self energy are finite.
Correspondingly $Z_2(\mu^2,\Lambda^2)$ stays finite when  the cutoff is sent 
to infinity, and we have $0 < Z_2(\mu^2,\Lambda^2) < 1$. Furthermore, in Landau 
gauge the ghost-gluon vertex is not ultraviolet divergent, and we can choose  
$\tilde{Z}_1=1$ \cite{Taylor:ff}. The Slavnov--Taylor identity for the 
quark-gluon vertex renormalization factor $Z_{1F}$ thus simplifies 
\cite{Marciano:1978su},
\begin{equation}
Z_{1F}=\frac{\tilde{Z}_1 Z_2}{\tilde{Z}_3} = \frac{Z_2}{\tilde{Z}_3}.
\end{equation}

Previous studies of the quark DSE in the so-called Abelian approximation  (see
e.g.\ the review \cite{Roberts:2000aa} or the recent summary \cite{Tandy:2003}
and references therein) as well as the recent investigation in
ref.~\cite{Bloch:2002eq} assume implicit cancellations between the full
quark-gluon vertex and the dressed gluon propagator in the integral over the
kernel of the quark DSE. Furthermore, in the tensor structure of
the quark-gluon vertex only a term proportional to $\gamma_\mu$ is employed.

In the current investigation we do not have to rely on implicit cancellations
since we calculated explicit solutions for the dressed gluon and ghost
propagators \cite{Fischer:2002hn}. We will also construct explicit
non-perturbative {\it ans\"atze} for the quark-gluon vertex. We note that very
recently lattice results for the quark-gluon vertex became available 
\cite{Skullerud:2002ge}. However, at present the error bars from such
simulations are too large to use the lattice results as guideline in the
construction of reliable {\it ans\"atze} for the quark-gluon vertex.
To proceed we assume that one may approximately factorize the quark-gluon 
vertex
\begin{equation}
\Gamma_\nu(q,k) = V_\nu^{abel}(p,q,k) \, W^{\neg abel}(p^2,q^2,k^2),
\label{vertex-ansatz}
\end{equation}
with $p$ and $q$ denoting the quark momenta and $k=(q-p)$ the gluon momentum. The
non-Abelian factor $W^{\neg abel}$ multiplies an Abelian part  $V_\nu^{abel}$,
which carries the tensor structure of the vertex. This  {\it ansatz} is
motivated by the aim to respect gauge invariance as much as  possible on the
present level of truncation.

The Slavnov-Taylor identity (STI) for the quark-gluon vertex is given by 
\cite{Marciano:1978su}
\begin{equation}
G^{-1}(k^2) \: i\: k_\mu \: \Gamma_\mu(q,k) = S^{-1}(p) \: H(q,p) - H(q,p) \: S^{-1}(q),
\label{quark-gluon-STI}
\end{equation}
with $G(k^2)$ being the ghost dressing function and $H(q,p)$ the ghost-quark 
scattering kernel.
At present the non-perturbative behavior of the ghost-quark scattering kernel
is unknown. Therefore we cannot solve this STI explicitly. However, comparing
the structure of eq.~(\ref{quark-gluon-STI}) with the corresponding Ward
identity of QED,
\begin{equation}
i \: k_\mu \: \Gamma_\mu^{QED}(q,k) = S^{-1}(p) -  S^{-1}(q),
\label{QED-quark-gluon-STI}
\end{equation}
we are able to infer some information: Whereas the ghost fields of QED decouple
from the theory and consequently do not show up in the Ward identity there is
an explicit factor of $G^{-1}(k^2)$ on the left hand side of
eq.~(\ref{quark-gluon-STI}). We therefore suspect the quark-gluon vertex of QCD
to contain an additional factor of $G(k^2)$ compared to the fermion-photon
vertex of QED. Some additional ghost dependent structure seems necessary to
account for the ghost-quark scattering kernel on the right hand side of
eq.~(\ref{quark-gluon-STI}). For simplicity we assume the whole ghost
dependence of the vertex to be contained in a non-Abelian factor multiplying an
Abelian tensor structure. For quenched calculations and in the context
of angular approximated DSEs a similar strategy has already been adopted in
refs.~\cite{Smekal,Ahlig}.
The Abelian part of the vertex, $V_\nu^{abel}$, will be adopted in the
following from analog {ans\"atze} in QED. The Ward identity
(\ref{QED-quark-gluon-STI}) has been solved {\it e.g.\/} in ref.\ 
\cite{Ball:1980ay} such that kinematical singularities are avoided.
Furthermore transverse parts of the fermion-photon vertex
have been fixed such that multiplicative renormalizability in the Abelian
fermion DSE is satisfied for all linear covariant gauges \cite{Curtis:1990zs}.
The resulting vertex is known as the Curtis-Pennington (CP) vertex.

The non-Abelian factor $W^{\neg abel}$ is chosen such that the resulting
quark propagator fulfills two conditions:
\begin{itemize}
\item[(i)]
The quark mass function $M(p^2)$ should be independent of the renormalization 
point $\mu^2$.
\item[(ii)] The anomalous dimension $\gamma_{m}$ of the mass function known 
from perturbation theory should be recovered in the ultraviolet.
\end{itemize}
In the course of this section we will prove the vertex {\it ansatz}
\beqa
W^{\neg abel}(p^2,q^2,k^2) &=&  G^2(k^2,\mu^2) \:\tilde{Z}_3(\mu^2,\Lambda^2)\: 
\frac{\left(G(k^2,\mu^2) \: 
\tilde{Z}_3(\mu^2,\Lambda^2) \right)^{-2d-d/\delta}}
{\left(Z(k^2,\mu^2) \: {Z}_3(\mu^2,\Lambda^2)\right)^d} 
\label{vertex_nonabel}\\
V^{abel}_\nu(p,q,k) &=& \Gamma_\nu^{CP}(p,q,k) \\ &=&
\frac{A(p^2,\mu^2)+A(q^2,\mu^2)}{2} \gamma_\nu + 
i \frac{B(p^2,\mu^2)-B(q^2,\mu^2)}{p^2-q^2} (p+q)_\nu
\nonumber\\
&& + \frac{A(p^2,\mu^2)-A(q^2,\mu^2)}{2(p^2-q^2)} (\pslash+\qslash)(p+q)_\nu
 \nonumber\\
&& + \frac{A(p^2,\mu^2)-A(q^2,\mu^2)}{2} \left[(p^2-q^2)\gamma_\nu - 
(\pslash-\qslash)(p+q)_\nu \right] \nonumber\\
&& \hspace*{3.8cm}  \times \frac{p^2+q^2}{(p^2-q^2)^2+(M^2(p^2)+M^2(q^2))^2}
\label{vertex_CP}
\eeqa
with the new parameter $d$ to satisfy the conditions (i) and (ii).
Again $k$ denotes the gluon momentum and $p$ and $q$ the quark momenta.
The anomalous dimension $\delta$ of the ghost propagator is
$\delta = -9 \:N_c/(44 \:N_c-8 \: N_f)$, {\it i.e.\/} the corresponding value 
at one loop order for $N_c$ colors and $N_f$ flavors.
The Abelian part of the vertex is given by the Curtis-Pennington (CP)
vertex $\Gamma_\nu^{CP}(p,q,k)$.

From a systematic point of view the newly introduced parameter $d$ in the
non-Abelian part of the vertex is completely arbitrary. Our numerical results,
however, will indicate that values around the somewhat natural choice $d=0$
match best with lattice simulations, see below.

For comparison we will also employ the much simpler vertex
$\Gamma_\nu = V^{abel}_\nu W^{\neg abel}$ with $W^{\neg abel}$ given by 
eq.~(\ref{vertex_nonabel}) and 
\beqa
V^{abel}_\nu(p,q,k) &=& Z_2(\mu^2,\Lambda^2) \, \gamma_\nu
\label{vertex_bare}
\eeqa
where we have taken the bare Abelian vertex, $\gamma_\nu$, multiplied with an
extra factor of $Z_2$. In Landau gauge this construction also satisfies the
conditions (i) and (ii), as will be shown in the next two
subsections. Furthermore in the numerical treatment we will additionally 
employ a vertex where the last, transverse term in Eq.~(\ref{vertex_CP}) is left out, 
{\it i.e.\/} a generalized Ball-Chiu (BC) vertex. In Landau gauge such a 
vertex also satisfies the conditions (i) and (ii).

\subsection{Multiplicative renormalizability of the quark equation \label{subsec-MR}}

To proceed we first substitute the simpler vertex {\it ansatz} 
(\ref{vertex_bare}) into the
quark equation (\ref{quark1}). By taking the Dirac trace once with and once
without multiplying the equation with $\pslash$ we project out the mass
function $M(p^2)$ and the vector self energy $A(p^2)$. 
To ease notation we will use in the following the abbreviations 
$x=p^2$, $y=q^2$ and $z=(p-q)^2$ for the squared momenta, also $s=\mu^2$ for 
the squared renormalization point and $L=\Lambda^2$ for the squared cutoff
of the theory. We arrive at
\beqa
M(x)A(x,s) &=& Z_2(s,L) \: m_0(L) + \frac{Z_2(s,L)}{3\pi^3} \int d^4q \Bigg\{
\frac{\alpha(z)}
{z \:(y+M^2(y))} \: Z_2(s,L) \: A^{-1}(y,s) \times  \nonumber\\
&&\hspace*{4.5cm}  \frac{\left(G(z,s) \: \tilde{Z}_3(s,L) \right)^{-2d-d/\delta}}
{\left(Z(z,s) \: {Z}_3(s,L)\right)^d} \: 3M(y) \Bigg\} \,,  \label{bare-dse-m}\\
A(x,s) &=& Z_2(s,L) + \frac{Z_2(s,L)}{3\pi^3} \int d^4q \Bigg\{
\frac{\alpha(z)}
{x \:z \:(y+M^2(y))} \: Z_2(s,L) \: A^{-1}(y,s) \times  \nonumber\\
&&\hspace*{1cm}  \frac{\left(G(z,s) \: \tilde{Z}_3(s,L) \right)^{-2d-d/\delta}}
{\left(Z(z,s) \: {Z}_3(s,L)\right)^d}
\left(-z + \frac{x+y}{2} + \frac{(x-y)^2}{2 \: z} \right) \Bigg\} \,, \nonumber\\
\label{bare-dse-a}
\eeqa
where we have used the following expression for the running coupling 
$\alpha$ in Landau gauge \cite{vonSmekal:1997is}
\beq
\alpha(x) = \frac{g^2}{4 \pi} \: Z(x,s)G^2(x,s) = 
\alpha(s) \: Z(x,s) \:G^2(x,s) \, .
\label{alphaDef}
\eeq
From eqs.~(\ref{bare-dse-m}) and (\ref{bare-dse-a}) it is clear
that the choice of the parameter $d=0$ is special since then the only input 
from the Yang--Mills sector is the running coupling.

The behavior of eqs.~(\ref{bare-dse-m}), (\ref{bare-dse-a}) under 
renormalization can be explored by changing the renormalization point 
$s=\mu^2$ to a new point $t=\nu^2$.
We first note that the factor stemming from the non-Abelian
part of the quark-gluon vertex is not affected by such a change:
\beqa
\frac{\left(G(z,s) \: \tilde{Z}_3(s,L) \right)^{-2d-d/\delta}}
{\left(Z(z,s) \: {Z}_3(s,L)\right)^d} =
\frac{\left(G(z,t) \: \tilde{Z}_3(t,L) \right)^{-2d-d/\delta}}
{\left(Z(z,t) \: {Z}_3(t,L)\right)^d} \,. \nonumber
\eeqa
This can be seen easily with the help of the relations
\beqa
G_0(x,L) &=& G(x,s) \: \tilde{Z}_3(s,L) \,, \\
Z_0(x,L) &=& Z(x,s) \: {Z}_3(s,L) \,,
\label{ZGren}
\eeqa
between the unrenormalized and renormalized ghost and gluon dressing functions.
Furthermore, the running coupling $\alpha(z)$ is independent of the
renormalization point, for a detailed discussion of this property using
the expression (\ref{alphaDef}) see ref.\ \cite{vonSmekal:1997is}.

From eq.~(\ref{Aren}) we infer
\beqa
Z_2(t,L) \: A^{-1}(x,t) = Z_2(s,L) \: A^{-1}(x,s) \,.
\eeqa
With the renormalization condition $A(t,t)=1$ we have
\beqa
Z_2(t,L) = Z_2(s,L)\: A^{-1}(t,s) \,,
\label{Z_2}
\eeqa
and subsequently
\beqa
A(x,t) = A(x,s) \:A^{-1}(t,s) \,.
\label{A_R}
\eeqa
Substituting eqs.~(\ref{Z_2}) and (\ref{A_R}) into the Dyson--Schwinger 
equations (\ref{bare-dse-m}) we find the mass function $M(x)$ to be 
independent of the renormalization point,
{\it i.e.\/} condition (i) is satisfied.
Note that without the extra factor of $Z_2$ in the Abelian part of the vertex
(\ref{vertex_bare}) we would violate this condition.

\begin{figure}
\centerline{
\epsfig{file=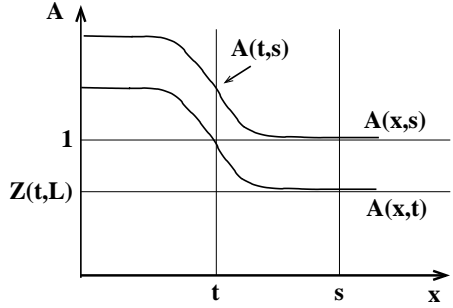,height=4cm}
}
\caption{\label{Afig} Sketch of a finite renormalization from a 
perturbative point $s$
to a non-perturbative point $t$ for the vector self-energy $A$.}
\end{figure}

Before we examine the case of the more sophisticated Curtis--Pennington type 
vertex (\ref{vertex_CP}) two remarks are in order. 
{\it First}, according to perturbation theory
we have $A(x \rightarrow \infty,s) \rightarrow 1$ and $Z_2(s,L) \rightarrow 1$ 
for large renormalization points $s$. 
However, this is just a special case of the general relation
\beqa
A(x \rightarrow \infty,s) \rightarrow Z_2(s,L) \label{A-limit}
\eeqa
which can be inferred from eqs.~(\ref{Z_2}) and (\ref{A_R}). In Fig.~\ref{Afig}
we sketch the vector self energy renormalized at two  different points $s$ and
$t$, with $s$ in the perturbative and $t$ in the non-perturbative region of
momentum. Of course, this mechanism will be found again in the numerical results,
see below.
{\it Second},  the appearance of the ghost and gluon renormalization factors
$Z_3(L)$ and $\tilde{Z}_3(L)$ in the interaction kernel of the quark equation
is due to the non-abelian part of the employed quark-gluon vertex {\it ansatz}.
Certainly, the renormalized functions $M(x)$ and $A(x,s)$ should not depend on
the cutoff of the integral. The balance of cutoff dependent quantities in the
equation is controlled by various factors of $Z_2(L)$ and $Z_m(L)$. We have to
take care not to disturb this balance by the vertex {\it ansatz}. Thus the
non-Abelian part of the quark-gluon vertex contains such powers of $Z_3(L)$ and
$\tilde{Z}_3(L)$ that the cutoff dependence of these quantities cancel. This
can be easily checked using the scaling behavior
\beqa
Z_3(s,L) &=& \left(\frac{\alpha(L)}{\alpha(s)}\right)^{\gamma} \,, \nonumber\\
\tilde{Z}_3(s,L) &=& \left(\frac{\alpha(L)}{\alpha(s)}\right)^{\delta} \,,
\label{Z-scaling}
\eeqa
of the renormalization factors for $L \rightarrow \infty$ and the relation
$\gamma+2\delta+1=0$ for the anomalous dimensions of the gluon and the ghost,
respectively.

Along the same lines as for the bare vertex construction we prove condition (i)
for the Curtis-Pennington type vertex. Plugging eqs.~(\ref{vertex_nonabel}),
(\ref{vertex_CP}) into the quark equation (\ref{quark1}) and projecting onto
$M(x)$ and $A(x)$ we arrive at
\beqa
M(x)A(x,s) &=& Z_2(s,L) m_0(L) + \frac{Z_2(s,L)}{3\pi^3} \int d^4q
\frac{\alpha(z)}
{z(y+M^2(y))}\frac{\left(G(z,s) \tilde{Z}_3(s,L) \right)^{-2d-d/\delta}}
{\left(Z(z,s) {Z}_3(s,L)\right)^d} \nonumber\\
&&\times A^{-1}(y,s) \: \left[\frac{3}{2}(A(x,s)+A(y,s))M(y) \right.\nonumber\\
&&\hspace*{2cm} + \frac{1}{2}(\Delta A\:M(y)-\Delta
B)\left(-z+2(x+y)-\frac{(x-y)^2}{z}\right)
\nonumber\\
&&\hspace*{2cm}\left.+\frac{3}{2}(A(x,s)-A(y,s))M(y)\Omega(x,y)(x-y)\right] 
\label{CP-M}
\eeqa
\beqa
A(x,s) &=& Z_2(s,L) + \frac{Z_2(s,L)}{3\pi^3} \int d^4q
\frac{\alpha(z)}
{xz(y+M^2(y))} \frac{\left(G(z,s) \tilde{Z}_3(s,L) \right)^{-2d-d/\delta}}
{\left(Z(z,s) {Z}_3(s,L)\right)^d} \nonumber\\
&&\times A^{-1}(y,s)\:\left[\left(-z+\frac{x+y}{2}+\frac{(x-y)^2}{2z}\right)
\frac{A(x,s)+A(y,s)}{2}\right.\nonumber\\
&&\hspace*{2cm}-\left(\frac{\Delta A}{2}
(x+y)+\Delta
B\:M(y)\right)\left(-\frac{z}{2}+(x+y)-\frac{(x-y)^2}{2z}\right)\nonumber\\
&&\hspace*{2cm}+\left.
\frac{3}{2}(A(x,s)-A(y,s))M(y)\Omega(x,y)\left(\frac{x^2-y^2}{2}-z\frac{x-y}{2}
\right)\right].
\nonumber\\
\label{CP-A}
\eeqa
Here we have used the abbreviations
\beqa
\Delta A &=& \frac{A(x,s)-A(y,s)}{x-y},\nonumber\\
\Delta B &=& \frac{B(x,s)-B(y,s)}{x-y},\nonumber\\
\Omega(x,y) &=& \frac{x+y}{(x-y)^2+(M^2(x)+M^2(y))^2}.\nonumber
\eeqa
With the help of the relations (\ref{Z_2}) and (\ref{A_R}) we find again the 
quark equations (\ref{CP-M}), (\ref{CP-A})
to be consistently renormalized. Employing
\beqa
B(x,t) = B(x,s) \, A^{-1}(t,s)
\eeqa
it can be seen directly that the mass function
\beqa
M(x)=B(x,s)/A(x,s)
\eeqa
is independent of the renormalization point.
(The same is true for a Ball--Chiu type vertex, which is the Curtis--Pennington
construction (\ref{vertex_CP}) without the transverse term proportional to
$\Omega(x,y)$. Note that in different gauges than Landau gauge only the
Curtis-Pennington construction would satisfy condition (i), similar to QED
\cite{Curtis:1990zs}.)

\subsection{Ultraviolet analysis of the quark equation 
\label{quark-UV-analysis}}

In this subsection we will show that the {\it ans\"atze} (\ref{vertex_CP}) and
(\ref{vertex_bare}) for the quark-gluon vertex both lead to the correct
perturbative limit of the quark mass function $M(x)$. We first examine the case
of the bare vertex construction, eq.~(\ref{bare-dse-m}).

The ghost and gluon dressing functions $G$ and $Z$ are slowly varying for large
momenta according to their perturbative limit. For loop momenta $y$ larger than
the external momentum $x$ we are therefore justified to employ the angular
approximation $G(z),Z(z) \rightarrow G(y),Z(y)$, see ref.\
\cite{Fischer:2002hn}. Furthermore, there is a region $x_0<y<x$ where the
approximation $G(z),Z(z) \rightarrow G(x),Z(x)$ is adequate. We are then able
to carry out the angular integrals in eq.~(\ref{bare-dse-m}).
If we additionally take the external momentum $x$ to be large enough
 all masses in the denominators become negligible since
the integral is dominated by loop momenta $y \approx x$. We then obtain
\beqa
M(x)A(x,s) &=& Z_2(s,L) m_0(L) \nonumber\\
&&\hspace*{0cm}+ \, \frac{Z_2(s,L)}{\pi} \frac{\alpha(x)}
{x}\: \int_{x_0}^x dy \: Z_2(s,L) \: A^{-1}(y,s) \:
\frac{\left(G(z,s) \tilde{Z}_3(s,L) \right)^{-2d-d/\delta}}
{\left(Z(z,s) {Z}_3(s,L)\right)^d} M(y) \nonumber\\
&&\hspace*{0cm}+ \, \frac{Z_2(s,L)}{\pi} \int_x^L dy
\frac{\alpha(y)}
{y} \: Z_2(s,L) \: A^{-1}(y,s)\:
\frac{\left(G(z,s) \tilde{Z}_3(s,L) \right)^{-2d-d/\delta}}
{\left(Z(z,s) {Z}_3(s,L)\right)^d} M(y) \,,\nonumber\\
\label{bare-dse-m-UV1}
\eeqa
where the integral from $y=0$ to $y=x_0$ has already been neglected.

For large momenta $y>x_0$ the wave function renormalization $A^{-1}$ and the
renormalization factor $Z_2$ cancel each other according to
eq.~(\ref{A-limit}). The ultraviolet limit of the ghost and gluon dressing 
functions from their respective DSEs in quenched approximation has been 
discussed in ref.~\cite{Fischer:2002hn} and found to be in agreement with
resumed perturbation theory to one-loop order. As will be seen in the
discussion beyond quenched approximation below (see eqs.~(\ref{anom_dim})),
we also obtain the correct anomalous dimensions for the ghost and 
gluon dressing functions $G$ and $Z$ in the case of $N_f \not=0$. We thus 
use the general perturbative limit 
\beqa
G(z) &=& G(s)\left[\omega\log\left(\frac{z}{s}\right)+1\right]^\delta \,, 
\nonumber\\
Z(z) &=& Z(s)\left[\omega\log\left(\frac{z}{s}\right)+1\right]^\gamma \,,
\label{Z_UV}
\eeqa
with $\omega=\beta_0\alpha(s)/(4 \pi)=(11N_c-2N_f)\alpha(s)/(12 \pi)$.
If we additionally substitute the scaling behavior of the renormalization
constants $Z_3$ and $\tilde{Z}_3$, eqs.~(\ref{Z-scaling}), and exploit the 
relation $\gamma+2\delta+1=0$ we arrive at
\beqa
M(x) &=&  m_0(L)
+ \frac{1}{\pi} \frac{\alpha(x)}{x} \int_{x_0}^x dy   M(y)
+ \frac{1}{\pi} \int_x^L dy \frac{\alpha(y)}{y}  M(y) \,.
\label{bare-dse-m-UV2}
\eeqa

This well known equation describes the ultraviolet behavior of the quark mass
function, see {\it e.g.\/} ref.\ \cite{Miransky:1985ib}. 
Employing the perturbative form of the running coupling,
\beqa
\alpha(y) = \alpha(s)\left[\omega\log\left(\frac{y}{s}\right)+1\right]^{-1} 
\label{alpha-UV}
\eeqa
one obtains in the chiral limit, $m_0(L) = 0$, 
the so-called {\it regular asymptotic} form
\beqa
M(x) = \frac{2 \pi^2 \gamma_m}{3}
\frac{-\langle \bar{\Psi}\Psi\rangle}
{x \left(\frac{1}{2} \ln(x/\Lambda^2_{QCD})\right)^{1-\gamma_m}} \,.
\label{chiral-M_UV}
\eeqa
Here $\langle \bar{\Psi}\Psi\rangle$ denotes the chiral condensate which is
discussed in more detail in the next subsection. In the case of
non-vanishing bare quark mass, $m_0(L) \ne 0$, the equation
(\ref{bare-dse-m-UV2}) is solved by the {\it irregular asymptotic} form,
\beqa
M(x) = M(s)\left[\omega\log\left(\frac{x}{s}\right)+1\right]^{-\gamma_m} \,.
\label{M_UV}
\eeqa
In this case we furthermore find
\beqa
\gamma_m &=& \frac{12}{11N_c-2N_f} \,, \\
m_0(L) &=& M(s) \left[\omega\log\left(\frac{L}{s}\right)+1\right]^{-\gamma_m} 
\label{m_0_UV}
\eeqa
in accordance with perturbation theory.

We thus have shown that the bare vertex construction (\ref{vertex_bare}) admits
a solution for the mass function $M(x)$ which has the correct perturbative
behavior for large momenta. A similar analysis is possible for the DSE with
the Curtis--Pennington type vertex, eq.~(\ref{CP-M}). As the vector self energy
goes to a constant in the limit of large momenta, eq.~(\ref{A-limit}), all
terms proportional to $A(x)-A(y)$ are suppressed in this limit. Furthermore,
according to the perturbative expression (\ref{M_UV}) the $\Delta B$-term
contributes at most subleading logarithmic corrections in eq.~(\ref{CP-M}). The
first term in the brackets reduces to the bare vertex form because $A(x,s)
\approx A(y,s)$ for large momenta $x,y$. Thus we obtain the same ultraviolet
limit from eq.~(\ref{CP-M}) as for the bare vertex construction. This is
certainly also the case if a Ball--Chiu type vertex is employed.

\section{The quark propagator in quenched QCD \label{quenched}}

In this section we will compare quark propagator results in quenched
approximation for three different
vertex types, which share the non-Abelian part proposed in
eq.~(\ref{vertex_nonabel}) but differ in their Abelian parts. We will employ
the bare vertex, eqs.~(\ref{vertex_bare}), and the Curtis-Pennington (CP) type
vertex, eqs.~(\ref{vertex_CP}). Furthermore we use a Ball-Chiu (BC) type
construction, which employs only the first three terms of the CP-vertex. In
Landau gauge all these vertex {\it ans\"atze} satisfy the conditions (i) and
(ii) formulated above eq.~(\ref{vertex_nonabel}). In order to compare the 
different vertex types on a quantitative level we will calculate the pion decay 
constant $f_\pi$ and the chiral condensate from the respective solutions for the 
quark mass function.

Before doing so we have to specify the effective quark interaction as input from
the Yang--Mills sector.

\subsection{Effective quark interaction}

Via the {\it ansatz} of the quark-gluon vertex the effective quark interaction
depends on the quark propagator functions $A(x,s)$ and $M(x)$ themselves. These
functions will be determined self-consistently in the process of the solution
the quark DSE. As further input the running coupling and, if $d\not=0$, the
gluon and the ghost propagator are needed. These will be taken from the results
of ref.\ \cite{Fischer:2002hn}.  

\begin{figure}
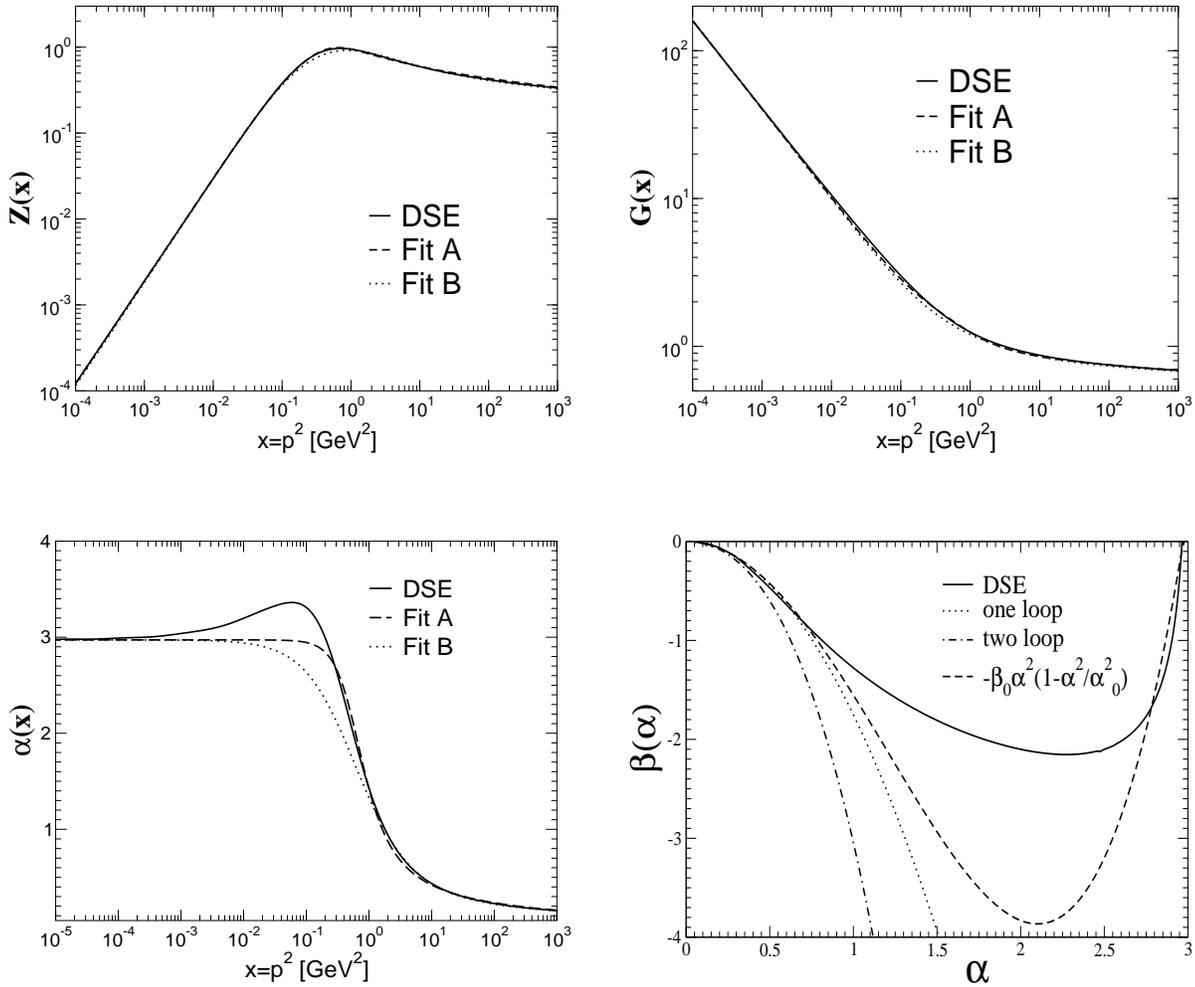

\vspace{0.5cm}
\centerline{
\epsfig{file=z.eps,width=7.5cm,height=6cm}
\hspace{0.5cm}
\epsfig{file=g.eps,width=7.5cm,height=6cm}
}
\vspace{1cm}
\centerline{
\epsfig{file=a.eps,width=7.5cm,height=6cm}
\hspace{0.5cm}
\epsfig{file=beta.eps,width=7.5cm,height=6cm}
}
\caption{\sf \label{new.dat}
Shown are the results for the gluon dressing function $Z$, the ghost dressing
function $G$ and the running coupling $\alpha$.
The two sets of fit functions are given in eqs.~(\ref{fitA}) and (\ref{fitB}).
The $\beta$-function corresponding to our DSE-solution is compared to 
the one- and two-loop expressions as well as to a polynomial in $\alpha$.}
\end{figure}

The DSEs of the Yang--Mills sector are scale-independent and the underlying
scale will be generated through dimensional transmutation during
renormalization. To translate this scale into physical units 
the scale $\Lambda_{QCD}$ is, in quenched approximation,
determined from the large-momentum behavior 
of the running coupling, or more precisely, by requiring $\alpha (M_Z^2) =
\alpha ((91.187 \, \mbox{GeV})^2)=0.118$. Technically this is achieved by 
fitting the functional form of the running coupling. 
We employ two different fit functions \cite{Alkofer:2002aa} for the running
coupling $\alpha(x)$:
\begin{eqnarray}
\mbox{Fit A:} \quad
\alpha(x) &=& \frac{\alpha(0)}{\ln[e+a_1 (x/\Lambda^2_{QCD})^{a_2}+
b_1(x/\Lambda^2_{QCD})^{b_2}]}\,,
\label{fitA}\\
\mbox{Fit B:} \quad
\alpha(x) &=& \frac{1}{a+(x/\Lambda^2_{QCD})^b}
\bigg[ a \: \alpha(0) + \nonumber\\
&&\hspace*{2cm} \left.\frac{4 \pi}{\beta_0} 
\left(\frac{1}{\ln(x/\Lambda^2_{QCD})}
- \frac{1}{x/\Lambda_{QCD}^2 -1}\right)(x/\Lambda^2_{QCD})^b \right]\,. 
\hspace*{0.5cm}
\label{fitB}
\end{eqnarray}
The value $\alpha(0)=8.915/N_c$ is known from the infrared analysis. In both
fits the ultraviolet behavior of the solution fixes the scale,  
$\Lambda_{QCD}=0.71\,\mbox{GeV}$. Especially the fit B is very robust leading 
to a change in $\Lambda_{QCD}$ of less than one percent when different fitting 
regions are chosen. Note that we have employed a MOM scheme, and thus 
$\Lambda_{QCD}$ has to be interpreted as $\Lambda_{MOM}^{N_f=0}$, {\it i.e.} 
this scale has the expected magnitude \footnote{For a discussion of the relation
of a $\widetilde{MOM}$-scheme to the $\overline{MS}$-scheme see section IV of 
ref.~\cite{Becirevic:1999hj}. Based on a three loop calculation the authors 
obtained the relation $\Lambda_{\overline{MS}} \simeq 0.346 
\Lambda_{\widetilde{MOM}}$. In our case this would result in the value 
$\Lambda_{\overline{MS}}=0.246$ which for zero flavors is even on the low side.}.
Fit A employs the four additional parameters:
$a_1=1.106$, $a_2=2.324$,
$b_1=0.004$, $b_2=3.169$.
Fit B has only two free parameters:
$a=1.020$, $b=1.052$.

In Landau gauge the gluon and ghost propagators are defined by
\beqa
D_{\mu \nu}(p) &=& 
%\delta^{ab} \left[ 
\left(\delta_{\mu \nu} - \frac{p_\mu p_\nu}{p^2}
\right) \frac{Z(p^2)}{p^2}  
%\right] 
\;, \label{Gl-prop}\\ 
D_{G}(p) &=& - 
%\delta^{ab} 
\frac {G(p^2)}{p^2} \;. \label{Gh-prop}
\eeqa
The gluon and ghost dressing functions, $Z(p^2)$ and $G(p^2)$, respectively,
are then described by
\begin{eqnarray}
R(x) &=& \frac{c \,(x/\Lambda^2_{QCD})^{\kappa}+d \,(x/\Lambda^2_{QCD})^{2\kappa}}
{1+ c \,(x/\Lambda^2_{QCD})^{\kappa}+d \,(x/\Lambda^2_{QCD})^{2\kappa}} \,, \nonumber\\
Z(x) &=& \left( \frac{\alpha(x)}{\alpha(\mu)} \right)^{1+2\delta} R^2(x) \,, \nonumber\\
G(x) &=& \left( \frac{\alpha(x)}{\alpha(\mu)} \right)^{-\delta} R^{-1}(x)  \,,
\label{zg-fit}
\end{eqnarray}
where $x=p^2$ and $c,d$ are fitting parameters for the auxiliary function $R(x)$.
They are given by $c=1.269$ and  $d=2.105$. Recall that
the anomalous dimension $\gamma$ of the gluon is related to the
anomalous dimension $\delta$ of the ghost by $\gamma=-1-2\delta$ and 
$\delta=-9/44$ for the number of flavors $N_f=0$.

Whereas Fit A is better in the region $0.3 \,\mbox{GeV}^2 < x<1 \,
\mbox{GeV}^2$ where $\alpha$ is strongly rising, Fit B is slightly better in
the region $1 \, \mbox{GeV}^2 < x < 10 \, \mbox{GeV}^2$. As can be
seen in Fig.~\ref{new.dat} both fits works very well and will be used as input
below.

\subsection{Pion decay constant, chiral condensate and quark masses 
\label{f-pi-sec}}

A correct calculation of the pion decay constants involves the pion
Bethe--Salpeter amplitudes including the subleading components, see  {\it
e.g.\/} refs.\ \cite{Langfeld:en,Maris:1997hd}. Apart from the
dressed quark propagator the Bethe--Salpeter equation  involves couplings
between quarks and gluons.  On the level of the quark DSE we have substituted
the full quark-gluon vertex by an vertex {\it ansatz}. However, at present it
is only known for certain cases how such a vertex {\it ansatz} in the quark DSE
translates to the corresponding quark-gluon coupling in the Bethe--Salpeter
equation \cite{Bender:2002as}. No method is known up to know to derive
the corresponding Bethe--Salpeter equation for dressed quark-gluon vertices as
the BC- or CP-vertex constructions.

We thus have to rely on the approximation \cite{Pagels:hd}
\beqa
f_\pi^2 &=& -\frac{N_c}{4 \pi^2} Z_2(s,L) \int dy \: y \: \frac{M(y) A^{-1}(y,s)}
{(y + M^2(y))^2} \left(M(y) - \frac{y}{2} \frac{dM(y)}{dy}\right) \,,
\label{f-pi-eq}
\eeqa
which incorporates only the effects of the leading pion Bethe-Salpeter
amplitude in the chiral limit \cite{Roberts:1994dr}. From a comparison of the
relative size of the amplitudes in model calculations
\cite{Maris:1997tm,Alkofer:2002bp} one concludes that the approximation
(\ref{f-pi-eq}) should lead to an underestimation of $f_\pi$ by 10-20 \%.

The {\it renormalization point independent} chiral condensate, $\langle
\bar{\Psi}\Psi\rangle$, can be extracted from the ultraviolet behavior of the
quark mass function in the chiral limit ({\it c.f.} eq.~(\ref{chiral-M_UV})):
\beqa
M(x) \, \stackrel{x \to L}{\longrightarrow} \, \frac{2 \pi^2 \gamma_m}{3}
\frac{-\langle \bar{\Psi}\Psi\rangle}{x \left(\frac{1}{2} 
\ln(x/\Lambda^2_{QCD})\right)^{1-\gamma_m}} \,.
\label{ch-fit}
\eeqa
Hereby $\Lambda_{QCD}$ is to be taken from a fit to the running coupling, 
{\it c.f.} eqs.~(\ref{fitA}), (\ref{fitB}).

The {\it renormalization point
dependent} chiral condensate $\langle \bar{\Psi}\Psi\rangle_\mu$ can be
calculated via \cite{Maris:1997hd} 
\beqa
-\langle \bar{\Psi}\Psi\rangle_\mu := Z_2(s,L) \, Z_m(s,L) \, N_c 
\,\mbox{tr}_D \int
\frac{d^4q}{(2\pi)^4} S_{ch}(q^2,s) \,,
\label{ch-cond}
\eeqa
where the trace is over Dirac indices, $S_{ch}$ is the quark propagator 
in the chiral limit
and the squared renormalization point is denoted by $s=\mu^2$.
To one-loop order both expressions for the condensate are connected by
\beq
\langle \bar{\Psi}\Psi\rangle_\mu = \left(\frac{1}{2}
\ln(\mu^2/\Lambda^2_{QCD})\right)^{\gamma_m}
\langle \bar{\Psi}\Psi\rangle \,,
\label{ch-loop}
\eeq
with $\gamma_m$ being the anomalous dimension  of the quark mass function.

For the calculation of the chiral condensate we first have to determine the
mass  renormalization constant $Z_m(s,L)$. Recall the formal structure of the
mass equation (\ref{bare-dse-m}), which is given as
\beqa
M(x)A(x,s) = Z_2(s,L) \,Z_m(s,L) \,m_R(s)  + Z_2(s,L) \, \Pi_M(x,s) \,,
\label{m_mu}
\eeqa
where $\Pi_M(x,s)$ represents the dressing loop. In order to extract 
$Z_m(s,L)$ from this equation
we have to clarify the meaning of $m_R(s)$ which is related to the 
unrenormalized mass by
\beqa
m_0(L)=Z_m(s,L) \,m_R(s),
\eeqa
{\it c.f.} eq.~(\ref{mass-ren}). Evaluating eq.~(\ref{m_mu}) at the
perturbative momentum $x=s$  the matter seems clear. We achieve consistency
with eqs.~(\ref{M_UV}) and (\ref{m_0_UV}), if
\beqa
m_R(s) &=& M(s) \label{mR} \,,\\
Z_m(s,L) &=& \left[\omega\log\left(\frac{L}{s}\right)+1\right]^{-\gamma_m} \,,
\eeqa
which is indeed the correct perturbative scaling of the renormalization 
constant $Z_m$ \cite{Roberts:2000aa}.

Certainly one could implicitly {\it define the finite parts} of $Z_m$ such that
the relation (\ref{mR}) holds in general for all renormalization points $s$.
Then the parameter $m_R$ in the renormalized QCD-Lagrangian would already know
about dynamical symmetry breaking. However, as the mass parameters of QCD are
supposed to be generated in the electroweak sector of the standard model one
could equally well argue that it is more systematic to exclude the effect of
mass generation by strong interaction from $m_R$.

In our numerical calculations we will choose $s$ to be sufficiently large,
therefore eq.~(\ref{mR}) is valid anyway. Then $Z_m$ is determined by
\beqa
Z_m(s,L) &=&  \frac{M(x)\, A(x,s) - Z_2(s,L) \, \Pi_M(x,s)}{Z_2(s,L)\, M(s)} 
\nonumber\\
&=& \frac{1}{Z_2(s,L)} - \frac{\Pi_M(s,s)}{M(s)} \,.
\eeqa
For the last equation we have set $x=s$ and have used the renormalization
condition $A(s,s)=1$.

In the numerical calculations we have to specify the masses $m_R(s)$ as input.
Choosing a perturbative renormalization point $s$ allows one to evolve the
masses $m_R(s)$ to a different scale $t$ by
\beqa
m_R(t) = m_R(s) \left(\frac{\ln(s/\Lambda_{QCD}^2)}
{\ln(t/\Lambda_{QCD}^2)}\right)^{\gamma_m}.
\eeqa
For $t=(2 \,\mbox{GeV})^2$ typical values for the masses of the light quarks 
are given by the Particle Data Group \cite{Hagiwara:2002pw}:
\beq
\frac{1}{2}(m_{u}+m_{d})(2 \,\mbox{GeV}) \approx 4.5 \mbox{MeV}, \hspace*{1cm}
m_{s}(2 \,\mbox{GeV}) \approx 100 \mbox{MeV}.
%\hspace*{1cm}
%m_{c}((m_c)^2) = 1 \mbox{GeV},
\label{masses-pdg}
\eeq
We will use similar masses in our calculations.

\subsection{Renormalization scheme and numerical method \label{ren-quenched}}

In the quark equation we employ a MOM regularization scheme
similar to the one used for the ghost and gluon equations in 
refs.\ \cite{vonSmekal:1997is,Fischer:2002hn}. For a fermion DSE this
technique has already been used 
in quenched $\mbox{QED}_4$ \cite{Kizilersu:2000qd}.
The formal structure of the quark equation is given by
\beqa
A(x,s) &=& Z_2(s,L) + Z_2(s,L) \, \Pi_A(x,s) \,, \\
M(x)A(x,s) &=& Z_2(s,L) \, Z_m(s,L) \, m_R(s)  + Z_2(s,L) \, \Pi_M(x,s) \,.
\label{m-formal}
\eeqa
We eliminate $Z_2$ from the first equation by isolating it on the left hand 
side and subtracting the same equation for $x=s$. With
\beqa
\frac{1}{Z_2(s,L)} = \frac{1}{A(x,s)} + \frac{1}{A(x,s)}\Pi_A(x,s)\,, 
\label{Z2-formal}
\eeqa
we then have
\beqa
\frac{1}{A(x,s)} = 1 - \frac{1}{A(x,s)}\Pi_A(x,s) + \Pi_A(s,s)\,, 
\label{A-formal}
\eeqa
using the renormalization condition $A(s,s)=1$.
In each iteration step we determine  $A(x)$ from eq.~(\ref{A-formal})
and subsequently $Z_2$ from eq.~(\ref{Z2-formal}). As a numerical check we 
determine $Z_2$ at different momenta $x=p^2$. In our calculations we find 
$Z_2$ to be independent of $p^2$ to a high accuracy. 
For the mass function $M(x)$ we use
\beqa
M(x)A(x,s) = Z_2(s,L) \, \Pi_M(x,s)
\eeqa
in the chiral limit and the subtracted equation
\beqa
M(x)A(x,s) = M(s) + Z_2(s,L)\, \Pi_M(x,s) - Z_2(s,L)\, \Pi_M(s,s)
\eeqa
if chiral symmetry is broken explicitly, {\it i.e.} $m_0 \not= 0$.

For the numerical iteration we employ a Newton method and represent the
dressing functions $A(x)$ and $M(x)$ with the help of Chebychev polynomials.
Furthermore, we use a numerical infrared cutoff $\epsilon$, which is taken
small enough for the numerical results to be independent of $\epsilon$.
Numerical difficulties arise in the case of the Curtis--Pennington type vertex
and even more for the Ball--Chiu construction. If the external momentum $x$ and
the loop momentum $y$ are both small and close to each other then the
derivative-like terms
\beqa
\Delta A = \frac{A(x,s)-A(y,s)}{x-y}, \hspace*{2cm} \Delta B = \frac{B(x,s)-B(y,s)}{x-y},
\eeqa
are hard to evaluate accurately. Although the functions $A(x)$ and $B(x)$  are
constant in the infrared and consequently should have derivatives close to zero
one encounters large values for $\Delta A$ and $\Delta B$ due to numerical
inaccuracies in $A$ and $B$. In order to evaluate $\Delta A$ and $\Delta B$
much more precisely  at small momenta we fit the expressions
\beqa
A(x,s) = \frac{A(0,s)}{1+a_1 \,(x/\Lambda_{QCD}^2)^{a_2}}\,, \hspace*{1cm}
B(x,s) = \frac{B(0,s)}{1+b_1 \,(x/\Lambda_{QCD}^2)^{b_2}}\,, \label{ab-fit}
\eeqa
with the parameters $a_1$, $a_2$, $b_1$ and $b_2$ to the numerically evaluated
functions. The scale $\Lambda_{QCD}=0.71 \,\mbox{GeV}$ has been determined
from the fits to the running coupling already in the last subsection.

For $x-y$ smaller than a suitable matching point we calculate the terms $\Delta
A$ and $\Delta B$ from the fits. This procedure eliminates the numerical errors
in the derivative terms and smoothes the numerical results considerably. In the
case of the Ball--Chiu type vertex the iteration process does not converge
unless we use these fits.

The renormalization condition employed in the ghost-gluon system of equations
is $G^2(s)Z(s)=1$ with $\alpha(s)=0.118$ at the squared renormalization point
$s=\mu^2$. Furthermore we choose a transversal tensor to contract the gluon
equation, {\it c.f.\/} ref.\ \cite{Fischer:2002hn}. The physical scale in the
quenched calculations is taken directly from the Yang--Mills results, {\it
i.e.\/} we use the experimental value $\alpha(M_Z^2)=0.118$ of the running
coupling at the mass of the Z-boson to fix the scale.

\subsection{Numerical results \label{quenched-num}}

\begin{figure}
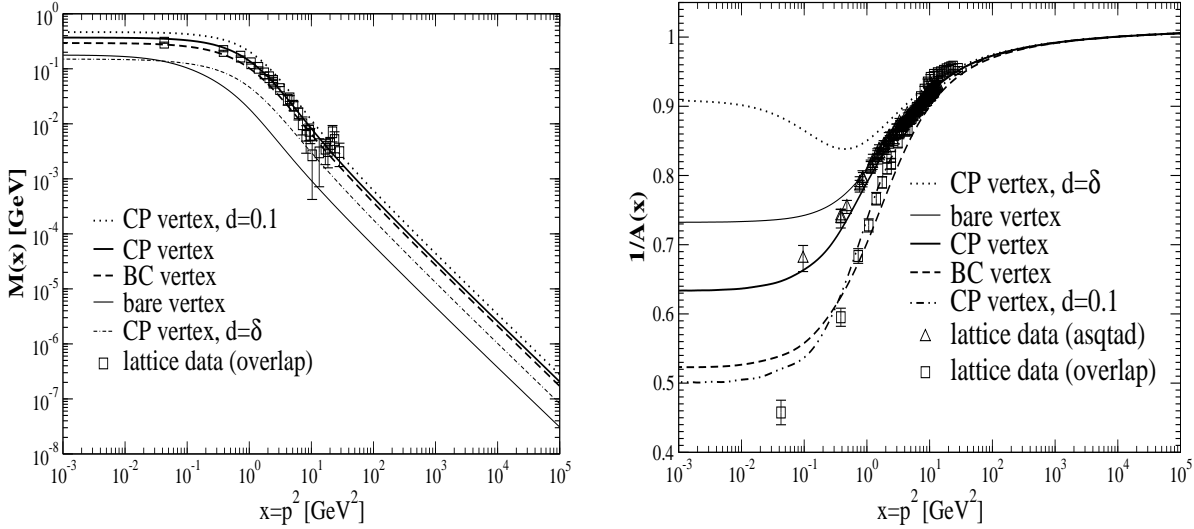

\vspace{0.7cm}
\centerline{
\epsfig{file=qM.vert.eps,width=7.5cm,height=7cm}
\hspace{0.5cm}
\epsfig{file=qA.vert.eps,width=7.5cm,height=7cm}
}
\caption{\label{qpic1.dat} 
The mass function $M(x)$ and the inverse vector self energy $1/A(x)$ of a
chiral quark are shown. We compare the results for five different vertices with
lattice data taken from refs.~\cite{Bonnet:2002ih,Bowman:2002kn}. } 
\end{figure}

In Fig.~\ref{qpic1.dat} we give our numerical solutions for the quark mass
function and the inverse vector self energy in the chiral limit, employing
'Fit A' for the effective quark interaction. We compare
results obtained with five different {\it ans\"atze} for the quark-gluon
vertex. For the generalized CP-vertex we investigate the 'natural' case $d=0$,
the value $d=\delta=-9/44$, already adopted in refs.~\cite{Smekal,Ahlig}, and
the value $d=0.1$. Furthermore we employed the bare vertex construction and a
Ball--Chiu type vertex. The corresponding masses at the momentum $p^2=0$, the
pion decay constant $f_\pi$, the renormalization point independent chiral
condensate and the fit parameters for the functions (\ref{ab-fit}) are
displayed in table \ref{qtable}.

\begin{table}
\begin{tabular}{c|c|c|c|c|c|c|c|c|c}
& M(0)& $f_\pi$ & $(-\langle \bar{\Psi}\Psi\rangle)^{1/3}$
&$(-\langle \bar{\Psi}\Psi\rangle)^{1/3}$
&$A^{-1}(0,M_Z^2)$&$a_1$&$a_2$&$b_1$&$b_2$\\
%&$a_1$&$a_2$&$b_1$&$b_2$ \\
%&&&\\
&\hspace{1mm}[MeV]\hspace{1mm}{ }&\hspace{1mm}[MeV]\hspace{1mm}{ }&\hspace{1mm}
[MeV] (calc.)\hspace{1mm}{ }&\hspace{1mm}[MeV] (fit)\hspace{1mm}{ }
%&$[\mbox{MeV}]^{-2a_2}$&&$[\mbox{MeV}]^{-2b_2}$\\
%&$A^{-1}(0,M_Z^2)$&$a_1$&$a_2$&$b_1$&$b_2$\\
&&&&\\
\hline
\mbox{bare vertex} &177&38.5&162&160&0.733&\hspace{1mm}3.05\hspace{1mm}{ }
&\hspace{1mm}0.99\hspace{1mm}{ }&
\hspace{1mm}0.06\hspace{1mm}{ }&\hspace{1mm}1.00\hspace{1mm}{ }\\
\mbox{CP d=$\delta$}&150&50.5&223&225&0.910&-&-&-&-\\
%\mbox{1BC vertex} &318&58.4&228&230&2.97&1.00&0.17&0.99\\
\mbox{BC-vertex} &293&62.6&276&284&0.523&1.10&0.99&0.29&0.92\\
\mbox{CP-vertex}&369&78.7&303&300&0.634&0.83&0.99&0.20&1.00\\
\mbox{CP d=0.1}&464&87.5&334&330&0.501&0.79&0.99&0.34&0.95
\end{tabular}
\caption{\label{qtable} 
The mass $M(0)$, the pion decay constant $f_\pi$ calculated with
eq.~(\ref{f-pi-eq}), the renormalization point independent chiral condensate
calculated with eqs.~(\ref{ch-cond}) and (\ref{ch-loop}), and the condensate
obtained by fitting the expression (\ref{ch-fit}) to the chiral mass function
in the ultraviolet for all five vertex types. Recall $\delta=-9/44$ in quenched
approximation. If not stated otherwise the parameter $d$ in the vertex
construction is taken to be $d=0$. For the case of the CP-vertex with
$d=\delta$ we did not get good fits in the infrared.}
\end{table}

The numerical results for the mass function all have a characteristic plateau
in the infrared and show the regular asymptotic behavior for large momenta,
{\it c.f.} eq.~(\ref{ch-fit}). The bare vertex construction and the CP type
vertex with $d=\delta$ both generate masses much smaller than typical
phenomenological values of $300-400$ MeV. The BC- and the CP-type construction
with $d=0$ provide good results, whereas the choice $d=0.1$ leads to a somewhat
large mass. The lattice calculations taken from refs.~\cite{Bonnet:2002ih}
(overlap fermions) and \cite{Bowman:2002kn} (improved staggered action, Asqtad)
favor masses around $300$ MeV with the caveat that they are obtained by an
extrapolation from sizeable bare quark masses to the chiral limit.
The numerical
solutions for the wave function renormalization $1/A$ can be seen in the right
diagram of Fig.~\ref{qpic1.dat}. Whereas the ultraviolet asymptotic behavior
of all vertex constructions is similar we observe sizeable differences for
small momenta. Again the bare vertex construction and the CP-vertex with
$d=\delta$ are clearly disfavored by the lattice data.

Our approximate calculation of the pion decay constant should underestimate the
experimental value $f_\pi= 93 \,\mbox{MeV}$ by 10-20 \%, {\it c.f.}
the discussion below eq.~(\ref{f-pi-eq}). We thus have best results for the
CP-vertex construction with $d=0$ and $d=0.1$. Furthermore we obtain very good
agreement between the two different methods to extract the chiral condensate. 
Compared to the phenomenological value 
%$(-\langle \bar{\Psi}\Psi\rangle)^{1/3} = (236 \pm 8) \, \mbox{MeV}$ 
%from QCD sum rules \cite{Leinweber:1997fn} or 
%$(-\langle \bar{\Psi}\Psi\rangle)^{1/3} = 227 \,\mbox{MeV}$ from the 
%phenomenological study summarized in ref.~\cite{Roberts:2000hi} 
$(-\langle \bar{\Psi}\Psi\rangle)^{1/3} \approx 250 \,\mbox{MeV}$
most of our results are larger.

Apart from the case $d=\delta$ we obtain very good fits for the scalar and 
vector self energy, $A(x)$ and $B(x)$, for small momenta. The results for the 
parameters $a_1, a_2, b_1, b_2$ in the fit functions given by 
eq.~(\ref{ab-fit}) can be found in table \ref{qtable}. It is interesting to 
note that the exponents $a_2$ and $b_2$ are found to be very close to one. 
Such a behavior could indicate a simple underlying functional form of the 
quark propagator. This will be explored in future work by a numerical 
continuation of our results to negative $p^2$, {\it i.e.} timelike momenta.

%%%%%%%%%%%%%%%%%%%%%%%%%%%%%%%%%%%%%%%%%%%%%%%%%%%%%%%%%%%%%%%%%%%%%%%%%%%%%%
\begin{figure}[t!]
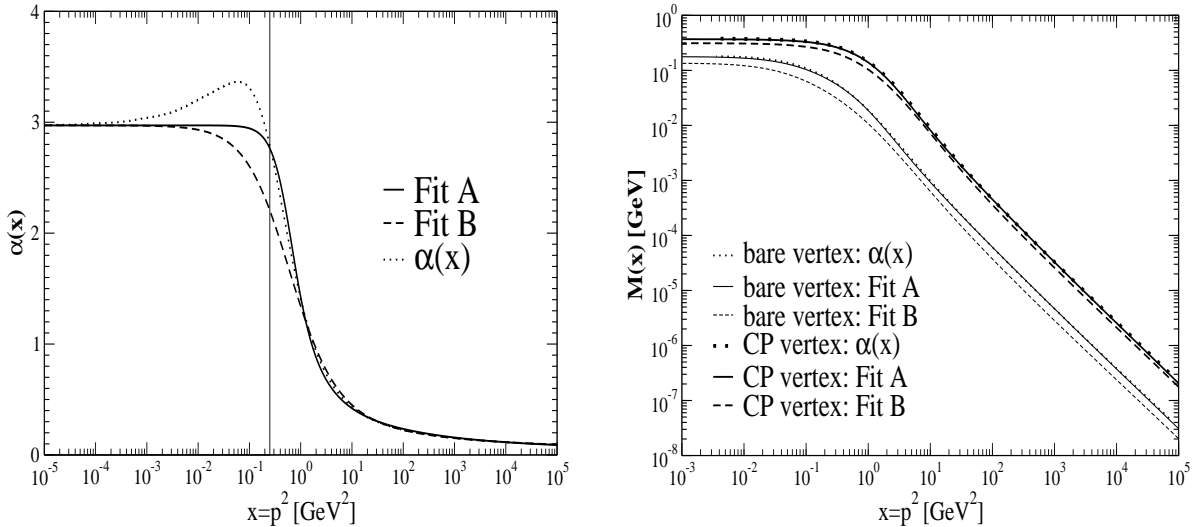

\vspace{1cm}
\centerline{
\epsfig{file=alpha.fit.eps,width=7.5cm,height=7cm}
\hspace{0.5cm}
\epsfig{file=qM.fit.eps,width=7.5cm,height=7cm}
}
\caption{\sf \label{qpic4.dat} Results for three different forms of the 
running coupling in the quark equation: The running coupling calculated in 
ref.~\cite{Fischer:2002hn} and the two fits given in eqs.~(\ref{fitA}), 
(\ref{fitB}). 
}
\end{figure}
%%%%%%%%%%%%%%%%%%%%%%%%%%%%%%%%%%%%%%%%%%%%%%%%%%%%%%%%%%%%%%%%%%%%%%%%%%%%
Fig.~\ref{qpic4.dat} compares results for the bare vertex and the CP-type 
construction with $d=0$ for three different forms of the running coupling in the 
interaction kernel of the quark equation. The two fit-functions, 'Fit A' 
and 'Fit B', have been given in eqs.~(\ref{fitA}), (\ref{fitB}). Furthermore 
we used the running coupling calculated from the quenched ghost and gluon DSEs
in ref.~\cite{Fischer:2002hn}. Although there is the (presumably) artificial 
bump at $p^2 \approx 0.1 \, \mbox{GeV}^2$ in the running coupling, the mass 
functions obtained from the DSE-result and from 'Fit A' are virtually 
indistinguishable. 'Fit B', however leads to somewhat smaller masses. This 
observation suggests that nearly all the dynamically generated mass is produced
from the integration strength above $p=500$ MeV, indicated by the vertical line 
in the plot of the running coupling. This is a favorable result as it would 
have been very unsatisfying if the artificial bump contributed a considerable 
amount to the quark mass function. 

%%%%%%%%%%%%%%%%%%%%%%%%%%%%%%%%%%%%%%%%%%%%%%%%%%%%%%%%%%%%%%%%%%%%%%%%%%%%%%%
\begin{figure}[t!]
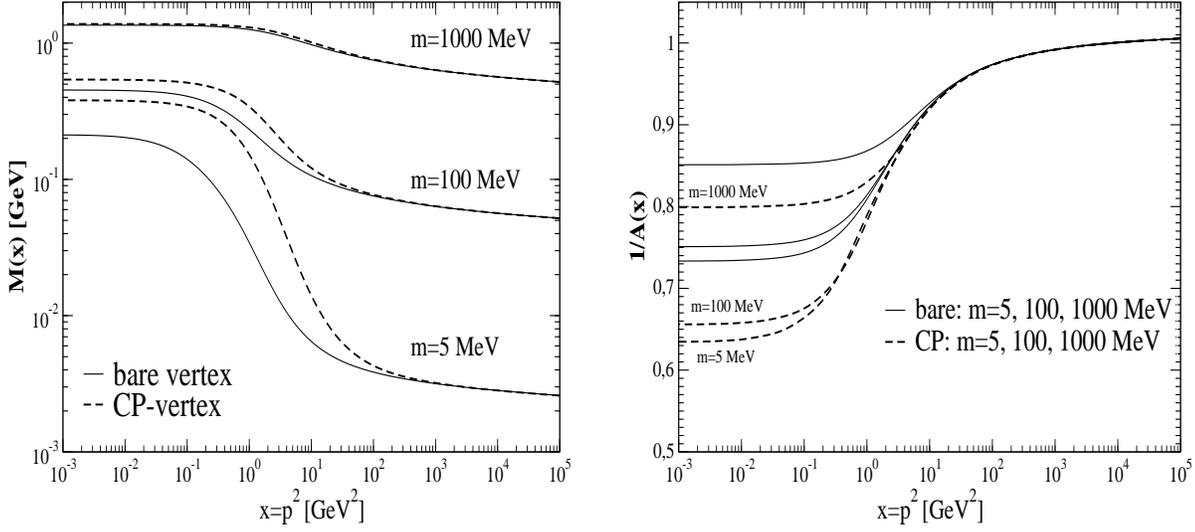

\vspace{1cm}
\centerline{
\epsfig{file=qM.mass.eps,width=7.5cm,height=7cm}
\hspace{0.5cm}
\epsfig{file=qA.mass.eps,width=7.5cm,height=7cm}
}
\caption{\sf \label{qpic5.dat} These diagrams show our results when three 
different bare quark masses are employed. In the diagram on the right small 
quark masses correspond to small values for $1/A$ in the infrared. 
}
\end{figure}
%%%%%%%%%%%%%%%%%%%%%%%%%%%%%%%%%%%%%%%%%%%%%%%%%%%%%%%%%%%%%%%%%%%%%%%%%%%%%%%
Finally we observe the effects of explicit chiral symmetry breaking in the plots 
of Fig.~\ref{qpic5.dat}. We give results for three different quark masses, 
$m(2 \,\mbox{GeV})=5$ MeV, $m(2 \,\mbox{GeV})=100$ MeV and $m(1 \,\mbox{GeV})=1000$
MeV. These values correspond roughly to the ones given by the Particle Data Group
for the up/down-quark, the strange-quark and the charm-quark \cite{Hagiwara:2002pw}.
For small momenta we note again that the dressed vertex generates more mass in 
the quark equation than the bare vertex construction. This effect becomes much 
less dominant for the heavy quarks, where more and more of the infrared mass stems 
from explicit chiral symmetry breaking and not from dynamical mass generation. 
Furthermore in accordance with the analytical determination we observe the same 
ultraviolet behavior of the mass function for both vertex constructions.

\section{The quark-loop in the gluon DSE}

In this section we focus on the inclusion of the back-reaction of the quarks on 
the ghost-gluon system, {\it i.e.} we will finally solve the complete set of 
mutually coupled Dyson-Schwinger equations for the quark, gluon and ghost 
propagator. To this end we incorporate the quark-loop in the truncation scheme 
for the gluon DSE which has been developed in ref.~\cite{Fischer:2002hn}. 

%%%%%%%%%%%%%%%%%%%%%%%%%%%%%%%%%%%%%%%%%%%%%%%%%%%%%%%%%%%%%%%%%%%%%%%%%%%%%%%%%
\begin{figure}[th!]
\vspace{0.5cm}
\centerline{
\epsfig{file=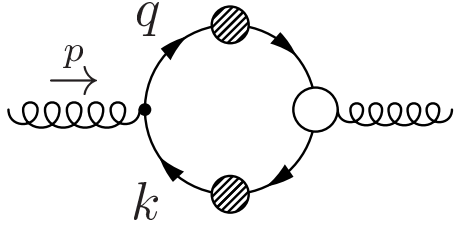,width=6.0cm,height=4cm}
}
\caption{\sf \label{quarkloop.fig} Diagrammatical representation of the quark-loop
in the Dyson-Schwinger equation for the gluon propagator.
}
\end{figure}
%%%%%%%%%%%%%%%%%%%%%%%%%%%%%%%%%%%%%%%%%%%%%%%%%%%%%%%%%%%%%%%%%%%%%%%%%%%%%%%%%%

The formal structure of the gluon equation is given by 
({\it c.f.} fig.~\ref{GluonGhostQuark})
\beq
\left[D(p)\right]^{-1}_{\mu \nu} = Z_3 \left[D^{(0)}(p)\right]^{-1}_{\mu \nu} + 
\Pi^{ghost}_{\mu \nu}(p)+\Pi^{gluon}_{\mu \nu}(p)+\Pi^{quark}_{\mu \nu}(p).
\eeq
The contributions from the ghost- and gluon-loop, $\Pi^{ghost}_{\mu \nu}(p)$ and
$\Pi^{gluon}_{\mu \nu}(p)$, are treated in detail in ref.~\cite{Fischer:2002hn}.
The contribution of the quark-loop to the gluon equation is given by 
\beq
 \Pi^{quark}_{\mu \nu}(p) = -\frac{g^2 N_f}{2 (2 \pi)^4} \,Z_{1F} \,\int d^4q \:\: 
 \mbox{Tr}\left\{ \gamma_\mu \,S(q)\, \Gamma_\nu(q,p) \,S(k)
 \right\},
\eeq
where $p$ is the external gluon momentum and $q$ and $k=(q-p)$ are the
momenta of the two quarks running in the loop, ({\it cf.} fig.~\ref{quarkloop.fig}). 
The trace is over Dirac indices. 

In eqs.~(\ref{vertex_nonabel}-\ref{vertex_CP}) we have proposed an effective 
quark-gluon vertex $\Gamma_\nu$ with Abelian and non-Abelian parts such that 
the quark equation is multiplicatively renormalizable and one-loop perturbation 
theory is recovered for large momenta. However, this construction is not capable 
to account as well for the one-loop behavior of the unquenched gluon equation 
unless we switch the momentum arguments of the non-Abelian part $W^{\neg abel}$, 
eq.~(\ref{vertex_nonabel}), from the gluon momentum to quark momenta. In the quark 
equation such a change of momentum arguments would either break Lorentz symmetry
by preferring one quark line of the quark-gluon vertex or changes the ultraviolet 
behavior of the quark equation. We therefore have to use different momentum 
assignments for the quark-loop and the quark equation. Certainly, this is a 
deficiency which has to be resolved by a more elaborate vertex construction in 
future work. The aim of the present study, however, is to present an effective 
construction which captures essential properties of the theory. 

Taking care of symmetries we propose the following {\it ansatz} for the non-Abelian 
part of the quark-gluon vertex in the quark-loop: 
\beqa
W^{\neg abel}_{quark-loop}(y,z,x) &=& G(y) G(z) \tilde{Z}_3(L) 
\frac{\left(G(y) \tilde{Z}_3(L) \right)^{-d-d/(2\delta)}}
{\left(Z(y) {Z}_3(L)\right)^{d/2}}
\frac{\left(G(z) \tilde{Z}_3(L) \right)^{-d-d/(2\delta)}}
{\left(Z(z) {Z}_3(L)\right)^{d/2}}. \nonumber\\
\eeqa
Here $x=p^2$ is the squared gluon momentum, $y=q^2$ and $z=k^2=(q-p)^2$ are the 
squared quark momenta, and $L=\Lambda^2$ is the squared cutoff. The Abelian part 
$V_\nu^{abel}$ of the vertex, given in eqs.~(\ref{vertex_CP},\ref{vertex_bare}), 
is symmetric with respect to the quark momenta as well.

Plugging the Curtis-Pennington type vertex into the quark-loop and contracting 
the free Lorenz-indices with the tensor ({\it c.f.} the treatment of the gluon 
DSE in ref.~\cite{Fischer:2002hn})
\beq
{\mathcal P}^{(\zeta )}_{\mu\nu} (p) = \delta_{\mu\nu} - 
\zeta \frac{p_\mu p_\nu}{p^2} 
\, , \label{tensor1} 
\eeq
we obtain
\setlength{\jot}{2mm}
\beqa
\Pi_{quark} &=&-\frac{g^2 N_f}{(2 \pi)^4} Z_2 \int d^4q \frac{G(y)}{y + M^2(y)} 
\frac{G(z)}{z + M^2(z)}  \frac{(G(y)\:G(z)\: \tilde{Z}_3^2(L))^{-d-d/(2\delta)}}
{ (Z(y)\:Z(z) \:Z_3^2(L))^{d/2}} \times \nonumber\\
&&\times A^{-2}(y)A^{-2}(z)\:\left\{\frac{A(y)+A(z)}{2} \left(W_1(x,y,z) A(y)A(z) + 
W_2(x,y,z) B(y)B(z)\right) \right. \nonumber\\
&&\hspace*{3cm}+ \frac{A(y)-A(z)}{2(y-z)} \left(W_3(x,y,z) A(y)A(z) + W_4(x,y,z) 
B(y)B(z) \right) \nonumber\\
&&\hspace*{3cm}+ \frac{B(y)-B(z)}{y-z} \left(W_5(x,y,z) A(y)B(z) + W_6(x,y,z) 
B(y)A(z) \right) \nonumber\\
&&\hspace*{-0.8cm}\left. + \frac{(A(y)-A(z)) (y+z)}{2((y-z)^2 + (M^2(y)+M^2(z))^2)} 
\left(W_7(x,y,z)A(y)A(z) + W_8(x,y,z)B(y)B(z) \right) \right\}, \nonumber\\  
\label{quarkloop}
\eeqa
with the kernels
\beqa
W_1(x,y,z) &=&\frac{\zeta z^2}{3 x^2} + z\left(\frac{2-\zeta}{3x} 
              - \frac{2\zeta y}{3x^2}\right)
              -\frac{2}{3}+\frac{(2-\zeta)y}{3x}  + \frac{\zeta y^2}{3x^2} \,, \\
W_2(x,y,z) &=& \frac{2 (4-\zeta)}{3x} \,, \\	       
W_3(x,y,z) &=& \frac{\zeta z^3}{3 x^2} - z^2\left(\frac{1+\zeta}{3x} 
              + \frac{\zeta y}{3x^2}\right) +z \left(\frac{1}{3} 
	      + \frac{(2\zeta-6)y}{3x}-\frac{\zeta y^2}{3x^2}  \right) 
	      \nonumber\\
	    &&+\frac{y}{3} 
	      - \frac{(\zeta+1)y^2}{3x} + \frac{\zeta y^3}{3x^2} \,, \\ 
W_4(x,y,z) &=& \frac{-2 \zeta z^2}{3x^2} + z \left(\frac{4}{3x}
              +\frac{4 \zeta y}{3x^2}  \right) 
              -\frac{2}{3} + \frac{4y}{3x} - \frac{2 \zeta y^2}{3 x^2}  \,,\\ 
W_5(x,y,z) &=&  \frac{\zeta z^2}{3x^2} -z\left( \frac{1+\zeta}{3x}
              +\frac{2\zeta y}{3x^2}\right)
              + \frac{1}{3} + \frac{(\zeta-3) y}{3x} 
	      + \frac{\zeta y^2}{3x^2}  \,,\\ 
W_6(x,y,z) &=& \frac{\zeta z^2}{3x^2} - z \left(\frac{3-\zeta}{3x}
              +\frac{2\zeta y}{3x^2} \right)
              + \frac{1}{3}  + \frac{(-\zeta-1) y}{3x}+ \frac{\zeta y^2}{3x^2}  \,,\\
W_7(x,y,z) &=&-\frac{z^2}{x} + z + \frac{y^2}{x} - y  \,,\\ 
W_8(x,y,z) &=& 2 \left( - \frac{z}{x} + \frac{y}{x} \right) \,.
\eeqa
Note that the symmetry factor $1/2$ and a factor $1/(3x)$ from the left hand 
side of the gluon equation have been absorbed in the kernels.  
From this expression the corresponding one for the bare vertex construction 
can be read off easily by setting $W_{3-8}=0$ and replacing the remaining 
factor $(A(y)+A(z))/2$ in eq.~(\ref{quarkloop}) by unity.

\subsection{Ultraviolet analysis of the quark-loop}

It is long known that the introduction of a cutoff $\Lambda$ in the gluon DSE 
results in artificial quadratic divergencies due to the violation of gauge 
invariance. Certainly, to recover the correct perturbative limit of the gluon 
propagator such terms have to be removed from the gluon equation by a suitable 
regularization procedure. Quadratic divergencies only occur in the part of the 
inverse gluon propagator proportional to $\delta_{\mu \nu}$. Therefore one way 
to eliminate the quadratic divergencies is to project onto the part proportional 
to $p_\mu p_\nu$ \cite{Brown:1988bm} by choosing $\zeta=4$ in the projection 
tensor (\ref{tensor1}).

Another unambiguous way is to subtract the quadratically divergent terms from 
the kernel by hand. This procedure is valid for general $\zeta$ and allows to 
estimate the influence of spurious longitudinal terms in the right hand side 
of the gluon equation on the solutions by varying the parameter $\zeta$. Certainly 
in a perfect truncation scheme the solution should be independent of $\zeta$. 
 
In ref.~\cite{Fischer:2002hn} it has been described in detail how to remove the
quadratic divergencies in the ghost- and gluon-loop. Therefore in the following 
we concentrate on the quark-loop. To identify the divergent terms we expand the 
dressing functions in the integrand of the quark-loop around large loop momenta 
$y$ with the difference $(z-y)$ still larger than any quark mass. 

To leading order this expansion amounts in the replacements 
\beqa
G(z) &\rightarrow& G(y), \nonumber\\
A(z) &\rightarrow& A(y), \nonumber\\
\frac{A(y) - A(z)}{y-z}  &\rightarrow& A^\prime(y), \nonumber\\
\frac{B(y) - B(z)}{y-z}  &\rightarrow& B^\prime(y), \nonumber\\
\frac{(A(y)-A(z)) (y+z)}{2((y-z)^2 + (M^2(y)+M^2(z))^2)} &\rightarrow& 
\frac{A^\prime(y) (y+z)}{2(y-z)},  
\eeqa
with the derivatives $A^\prime$ and $B^\prime$.
Note that the first two equations are identical to the angular approximation 
employed previously in the ultraviolet analysis of the quark equation. 
For large momenta $x$ and $z$ the denominators in eq.~(\ref{quarkloop})
simplify and the angular integrals are trivially performed using the integrals 
given in appendix A. We arrive at 
\setlength{\jot}{3mm}
\beqa
 \Pi_{quark}^{UV} &=&-\frac{g^2 N_f}{16 \pi^2} Z_2 \int dy \:G^2(y)\: 
 \frac{G(y)^{-2d-d/\delta}}{ Z(y)^{d}} \: A^{-2}(y) \times \nonumber\\
 && \hspace*{1.9cm}
\left\{A(y) \left(\frac{-2}{3y}+\frac{4-\zeta}{3x}
 + \frac{2 (4-\zeta)}{3xy} M^2(y)\right)
\right. \nonumber\\
&& \hspace*{1.9cm}+ \frac{A^\prime(y)}{2} \left(\frac{1}{3}+
  \frac{-2(4-\zeta)y}{3x} + \left(\frac{-2}{3y} + \frac{2(4-\zeta)}{3x} 
  \right) M^2(y) \right) \nonumber\\
&& \hspace*{1.9cm}+ B^\prime(y) M(y) \left( \frac{2}{3y} 
        - \frac{4-\zeta}{3x} \right)
\left. + \frac{A^\prime(y)}{2} 
\left(\frac{4y}{x}-1 + \frac{4}{x}M^2(y) \right) \right\}. \hspace*{1.1cm}  
\label{UV_quarkloop}
\eeqa
Keeping in mind a factor $(1/y)$ hiding in the derivatives we identify three 
quadratically divergent terms: $(4-\zeta)/3x$ in the second line, 
$-2(4-\zeta)y/3x$ in the third line and $4 y/x$ in the last line. The first 
two of them are artefacts of the regularization and will be subtracted from 
the kernels. However, we encounter the additional $\zeta$-independent 
quadratic divergent term $4 y/x$ originating from the transverse part of the 
Curtis-Pennington vertex. Such a term is already known from corresponding 
studies in QED \cite{Bloch}. Although first suggestions have been made how 
the Curtis-Pennington vertex should be modified to avoid this problem 
\cite{Pennington:1998cj}, a convincing solution has not been found yet. In 
our study we therefore choose the pragmatic strategy of subtracting 
this term by hand together with the other quadratically divergent parts.

Moreover we subtract all further terms proportional to $(4-\zeta)$. Although 
these terms are not quadratically divergent they are artefacts of the 
regularization. We then obtain a $\zeta$-independent expression for the quark 
loop at large momenta. In ref.~\cite{Fischer:2002hn} similar $\zeta$-independent 
expressions for the ghost- and gluon-loop have been derived. We therefore 
arrive at a transversal right hand side of the gluon equation for large 
momenta as required in Landau gauge. 

Collecting all modifications together we have the new kernels
\setlength{\jot}{2mm}
\beqa
\widetilde{W}_1(x,y,z) &=& W_1(x,y,z) - \frac{(y+z)(4-\zeta)}{6x}, \\
\widetilde{W}_2(x,y,z) &=& 0, \\ 
\widetilde{W}_3(x,y,z) &=& W_3(x,y,z) + \frac{2zy(4-\zeta)}{3x}, \\ 
\widetilde{W}_4(x,y,z) &=& W_4(x,y,z) - \frac{(y+z)(4-\zeta)}{3x}, \\
\widetilde{W}_5(x,y,z) &=& W_5(x,y,z) - \frac{(y+z)(4-\zeta)}{6x}, \\
\widetilde{W}_6(x,y,z) &=& W_6(x,y,z) - \frac{(y+z)(4-\zeta)}{6x}, \\
\widetilde{W}_7(x,y,z) &=& W_7(x,y,z) - \frac{(y-z)(y+z)}{x},\\
\widetilde{W}_8(x,y,z) &=& W_8(x,y,z).
\eeqa   
Note that the subtracted terms are chosen to preserve the symmetry of the 
kernels with respect to the squared quark momenta $y$ and $z$.

Without quadratic divergences we are in a position to extract the leading 
logarithmic divergence of the quark-loop. With modified kernels the ultraviolet 
limit of the quark-loop is given by
\beqa
 \Pi_{quark}^{UV} &=&-\frac{g^2 N_f}{16 \pi^2} Z_2 \int dy \:G^2(y)\: 
 \frac{G(y)^{-2d-d/\delta}}{ Z(y)^{d}} \: A^{-2}(y) \times \nonumber\\
 && \hspace{2.5cm}
\left\{A(y) \frac{-2}{3y} 
   + \frac{A^\prime(y)}{2} \left(\frac{1}{3}
  + \frac{-2}{3y} M^2(y)\right)  \right.\nonumber\\
&& \hspace{2.5cm}+ B^\prime(y) M(y)  \frac{2}{3y}
\left. + \frac{A^\prime(y)}{2} 
\left(-1+\frac{4}{x}M^2(y)\right)  \right\}.  
\label{UV2_quarkloop}
\eeqa
Similar to the situation in the DSE for the quark mass function the leading 
ultraviolet term is the first term in the curly brackets. For the ghost and 
gluon dressing functions we employ the perturbative {\it ansatz}
\beqa
G(z) &=& G(s)\left[\omega\log\left(\frac{z}{s}\right)+1\right]^\delta \,, 
\nonumber\\
Z(z) &=& Z(s)\left[\omega\log\left(\frac{z}{s}\right)+1\right]^\gamma \,,
\eeqa
and determine the anomalous dimensions $\delta$ and $\gamma$ as well as the 
coefficient $\omega$ selfconsistently as follows. Substituting the ultraviolet 
limit of the vector self energy, eq.~(\ref{A-limit}), and choosing the 
perturbative renormalization condition $G(s)=Z(s)=1$ we arrive at
\beq
\Pi_{quark}^{UV}(p) = \frac{2 N_f}{3 (2\delta+1) \omega} \frac{g^2}{16 \pi^2} 
\left\{\left[ \omega \log\left(\frac{L}{s}\right)+1 \right]^{2\delta+1} -
\left[ \omega \log\left(\frac{x}{s}\right)+1 \right]^{2\delta+1} \right\}. 
\label{quark-anom}
\eeq
Combining this expression with the results for the ghost- and gluon-loop from 
ref.~\cite{Fischer:2002hn} we obtain as ultraviolet limit of the gluon equation 
\beqa
\left[ \omega \log\left(\frac{x}{s}\right)+1 \right]^{-\gamma} &=&
Z_3 + \left(\frac{N_c g^2}{96 \pi^2 \omega (2\delta+1)} 
- \frac{7 N_c g^2}{48 \pi^2 \omega (2\delta+1)} 
+ \frac{2 N_f g^2}{48 \pi^2 \omega (2\delta+1)}\right)\times \nonumber \\
&&\hspace*{0.8cm}\left\{
\left[ \omega \log\left(\frac{L}{s}\right)+1 \right]^{2\delta+1} -
\left[ \omega \log\left(\frac{x}{s}\right)+1 \right]^{2\delta+1} \right\} \,. 
\label{gluoneq_uv}
\eeqa
The corresponding expression for the ghost equation reads \cite{Fischer:2002hn}
\beqa
\left[ \omega \log\left(\frac{x}{s}\right)+1 \right]^{-\delta} &=&
\tilde{Z}_3 - \frac{3 N_c g^2}{64 \pi^2 \omega(\gamma+\delta+1)} 
\times \nonumber\\
&&\hspace*{0.8cm}\left\{
\left[ \omega \log\left(\frac{L}{s}\right)+1 \right]^{\gamma+\delta+1} - 
\left[ \omega \log\left(\frac{x}{s}\right)+1 \right]^{\gamma+\delta+1} \right\}\,. 
\hspace*{0.9cm} 
\eeqa
The renormalization constants $Z_3(s,L)$ and $\tilde{Z}_3(s,L)$ cancel the 
cutoff dependence, {\it i.e.} the respective first terms in the curly brackets. 
Thus, the power and the prefactor of the second term have to match with the left 
hand side of the equations. This leads to the anomalous dimensions 
\beqa
\gamma &=& \frac{-13 N_c + 4 N_f}{22 N_c - 4 N_f} \nonumber\\
\delta &=& \frac{-9 N_c}{44 N_c - 8 N_f}
\label{anom_dim}
\eeqa
which are related by $\gamma+2\delta+1=0$ and in accordance with one-loop 
perturbation theory for arbitrary numbers of colors $N_c$ and flavors $N_f$.
For the coefficient $\omega$ one obtains
\beq
\omega = (11N_c-2N_f)\alpha(s)/(12 \pi) = \beta_0\alpha(s)/(4 \pi).
\eeq
When combined according to eq.~(\ref{alphaDef}) our ghost and gluon dressing 
functions lead to the correct one-loop running of the coupling $\alpha(x)$ at 
large momenta.
\setlength{\jot}{0mm}

\subsection{Infrared analysis of the quark-loop \label{ir-quark}}

The infrared analysis of the ghost and gluon DSEs in a truncation with bare 
ghost-gluon vertex has been performed in 
refs.~\cite{Zwanziger:2001kw,Lerche:2002ep,Fischer:2002hn}. To leading order 
the power law {\it ansatz}
\beq
Z(x) \sim x^{2\kappa} , \hspace*{1cm} G(x) \sim x^{-\kappa}, \label{ir-ansatz}
\eeq
for the ghost and gluon dressing functions at small momenta $x=p^2$ has been 
employed. For a transverse projection tensor (\ref{tensor1}), {\it i.e.} 
$\zeta=1$, one obtains $\kappa= (93-\sqrt{1201})/98 \approx 0.5954$ for 
the exponent of the dressing functions and subsequently the fixed point 
$\alpha(0) \approx  8.915/N_c$ for the running coupling in the infrared.

These results have been obtained in a truncation where the ghost-loop dominates 
the gluon-loop in the infrared. Therefore in order to investigate the effects 
of dynamical quarks in the system we compare the infrared behavior of the 
quark-loop with the one of the ghost-loop. Substituting the {\it ansatz} 
(\ref{ir-ansatz}) into the gluon equation and calculating the ghost-loop along 
the lines of the infrared analysis given in 
refs.~\cite{Lerche:2002ep,Fischer:2002hn} one finds the ghost-loop to be 
proportional to $x^{-2\kappa}$. For the quark-loop including the effects of dynamically
generated quark masses we obtain 
\beq
 \Pi_{quark}^{IR}(p) \sim x^{-2\kappa+2+\kappa d/\delta}.
\eeq  
Therefore {\it the quark-loop is suppressed for small momenta} provided the 
parameter $d$ fulfills the condition
\beq
d < \frac{-2\delta}{\kappa}.
\eeq
As we have $\kappa \approx 0.5954$ and $\delta=-1/4$ for $N_c=3$ and $N_f=3$ we 
find the condition $d<0.84$, which is satisfied for all quark-gluon vertices 
employed in our calculation. From a numerical point of view we encounter serious 
instabilities in the quark equation once $d$ is taken to be larger than 
$d \approx 0.2$. 

We conclude that with dynamically generated quark masses the quark-loop does not 
change the infrared behavior of the ghost 
and gluon dressing functions found in refs.~\cite{Zwanziger:2001kw,Lerche:2002ep}. 
In pure Yang-Mills theory as well as in QCD we thus have an infrared finite or 
vanishing gluon propagator and a ghost propagator which is more divergent than a 
simple pole. The Kugo-Ojima confinement criterion \cite{Kugo:1979gm, Kugo:1995km} 
and Zwanziger's horizon condition \cite{Zwanziger:1992ac, Zwanziger:2001kw} are 
both fulfilled not only in pure Yang-Mills theory (see {\it e.g.} 
refs.~\cite{Alkofer:2002aa}) but also in QCD. 

The reasoning above reveals a selfconsistent picture valid for a small number of 
flavours: As has been demonstrated in our quenched calculations the combined 
strength of the ghost and gluon propagator generates sizeable dynamical quark 
masses in the quark DSE. In the unquenched case these masses suppress the quark 
loop in the infrared such that the ghost and gluon propagators are hardly changed
in the infrared
and in turn nearly the same amount of mass is generated in the quark equation as
in the quenched case. This scenario is verified by our numerical calculations 
presented in the next section. On the other hand for a sufficiently large number
of light flavours we expect a different selfconsistent picture to apply: Chiral symmetry
should be restored. With vanishing quark masses the quark loop will contribute to 
the gluon DSE at small momenta and the infrared behaviour of the Yang-Mills sector 
will be changed. Especially it is expected that the value of the fixed point of the 
running coupling is decreased dramatically. This in turn drives the quark equation 
to the chirally symmetric solution with $M(p^2) \equiv 0$. This second selfconsistent 
scenario as well as the situation under the presence of a small amount of explicit 
chiral symetry breaking are subject to future investigations.

\section{Numerical results \label{res-unquenched}}

The numerical treatment of the integrals in the ghost and gluon equations has been 
described in detail in ref.~\cite{Fischer:2002hn}. The iteration process is done for 
the ghost-gluon system and the quark equations separately: we first iterate the $N_f$ 
mutually uncoupled quark systems until convergence is achieved, feed the output into 
the ghost and gluon system, iterate until the ghost-gluon system converges, feed the 
output back into the quark equations and so on, until complete convergence of all 
equations is achieved. We renormalized at the point $s=\mu^2$ given by 
$\alpha(s)=0.2$ and used a transverse tensor to contract the gluon 
equation, $\zeta=1$.

In contrast to the quenched calculation we fix the physical scale of the system not 
by the condition $\alpha(M_Z^2)=0.118$ but by adjusting the pion decay constant. 
This choice has an important advantage: Whereas the behavior of the running 
coupling for large momenta depends strongly on $N_f$ the pion decay constant turns 
out to be almost independent of the number of flavors. In our quenched calculation
we obtained $f_\pi=78.7$ MeV for the CP-type vertex with $d=0$. Recalling that our
approximation (\ref{f-pi-eq}) should lead to an underestimation of $f_\pi$ by
about 10-20 \% this value is in good accordance with experiment as 
$f^{exp.}_\pi = 93$ MeV. We therefore chose the scale in the unquenched 
calculations to lead to the same decay constant for the CP-type vertex with $d=0$. 

%%%%%%%%%%%%%%%%%%%%%%%%%%%%%%%%%%%%%%%%%%%%%%%%%%%%%%%%%%%%%%%%%%%%%%%%%%%%%%%%%%%%%
\begin{table}
\begin{tabular}{c||c|c||c|c||c|c||c|c||c|c}
&\multicolumn{2}{c||}{M(0)} & \multicolumn{2}{c||}{$f_\pi$} &  
\multicolumn{2}{c||}{$(-\langle \bar{\Psi}\Psi\rangle)^{1/3}$}&
\multicolumn{2}{c||}{$\alpha(M_Z)$}&\multicolumn{2}{c}{$\Lambda_{QCD}^{MOM}$}\\ 
& \multicolumn{2}{c||}{[MeV]} & \multicolumn{2}{c||}{[MeV]} &  
\multicolumn{2}{c||}{[MeV]}&
\multicolumn{2}{c||}{-}&\multicolumn{2}{c}{[MeV]}\\ \hline vertex 
&\hspace{2mm}qu.\hspace{2mm}{ }&\hspace{2mm}unqu.\hspace{2mm}{ }
&\hspace{2mm}qu.\hspace{2mm}{ }&\hspace{2mm}unqu.\hspace{2mm}{ }
&\hspace{2mm}qu.\hspace{2mm}{ }&\hspace{2mm}unqu.\hspace{2mm}{ }
&\hspace{2mm}qu.\hspace{2mm}{ }&\hspace{2mm}unqu.\hspace{2mm}{ }
&\hspace{2mm}qu.\hspace{2mm}{ }&\hspace{2mm}unqu.\hspace{2mm}{ }\\ \hline
\mbox{bare d=0}     &177 &176 &38.5 &38.4 &160 &170 &0.118 &0.146 &710 &748\\
\mbox{CP d=$\delta$}&150 &133 &50.5 &46.3 &225 &230 &0.118 &0.146 &710 &746\\
\mbox{CP d=0}       &369 &360 &78.7 &78.7 &300 &310 &0.118 &0.143 &710 &672\\
\mbox{CP d=0.1}     &464 &437 &87.5 &85.5 &330 &330 &0.118 &0.142 &710 &640
\end{tabular}
\caption{\sf \label{unqtable} A comparison between the quenched (qu.) and unquenched 
(unqu.) results for the quark mass M(0), the pion decay constant $f_\pi$, the 
renormalization point independent chiral condensate, the running coupling at the mass 
of the Z-boson and $\Lambda_{QCD}^{MOM}$ for different vertices and values of the  
parameter $d$. The unquenched calculations are done for $N_f=3$ chiral quarks. 
Furthermore we have $\delta=-9N_c/(44N_c-8N_f)=-0.25$ in the present case.}
\end{table}
%%%%%%%%%%%%%%%%%%%%%%%%%%%%%%%%%%%%%%%%%%%%%%%%%%%%%%%%%%%%%%%%%%%%%%%%%%%%%%%%%%%%%
\begin{figure}[th!]
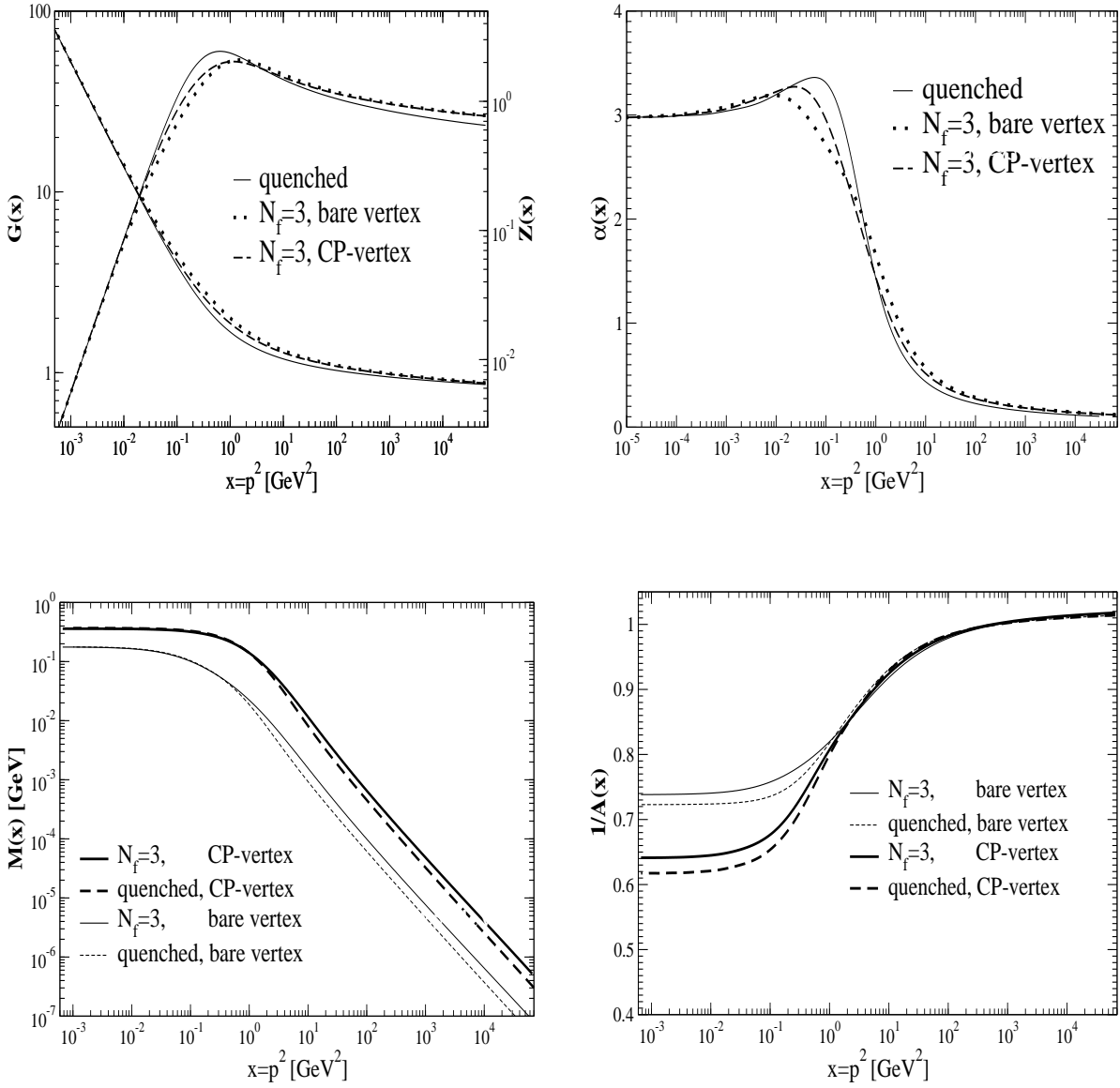

\vspace{0.5cm}
\centerline{
\epsfig{file=zg.1.eps,width=7.5cm,height=7cm}
\hspace{0.5cm}
\epsfig{file=alpha.1.eps,width=7.5cm,height=7cm}
}
\vspace{1.3cm}
\centerline{
\epsfig{file=M.1.eps,width=7.5cm,height=7cm}
\hspace{0.5cm}
\epsfig{file=1A.1.eps,width=7.5cm,height=7cm}
}
\caption{\sf \label{pic1.dat} Displayed are the ghost and gluon dressing function, 
$Z$ and $G$, the running coupling $\alpha$, the quark mass function $M$ and the 
inverse vector self energy $1/A$. The calculations are done quenched and unquenched 
with $N_f=3$ quarks in the chiral limit. The parameter $d$ in the vertices is set to 
$d=0$. 
}
\end{figure}
%%%%%%%%%%%%%%%%%%%%%%%%%%%%%%%%%%%%%%%%%%%%%%%%%%%%%%%%%%%%%%%%%%%%%%%%%%%%%%%%%%%%
%%%%%%%%%%%%%%%%%%%%%%%%%%%%%%%%%%%%%%%%%%%%%%%%%%%%%%%%%%%%%%%%%%%%%%%%%%%%%%%%%%%%
\begin{figure}[th!]
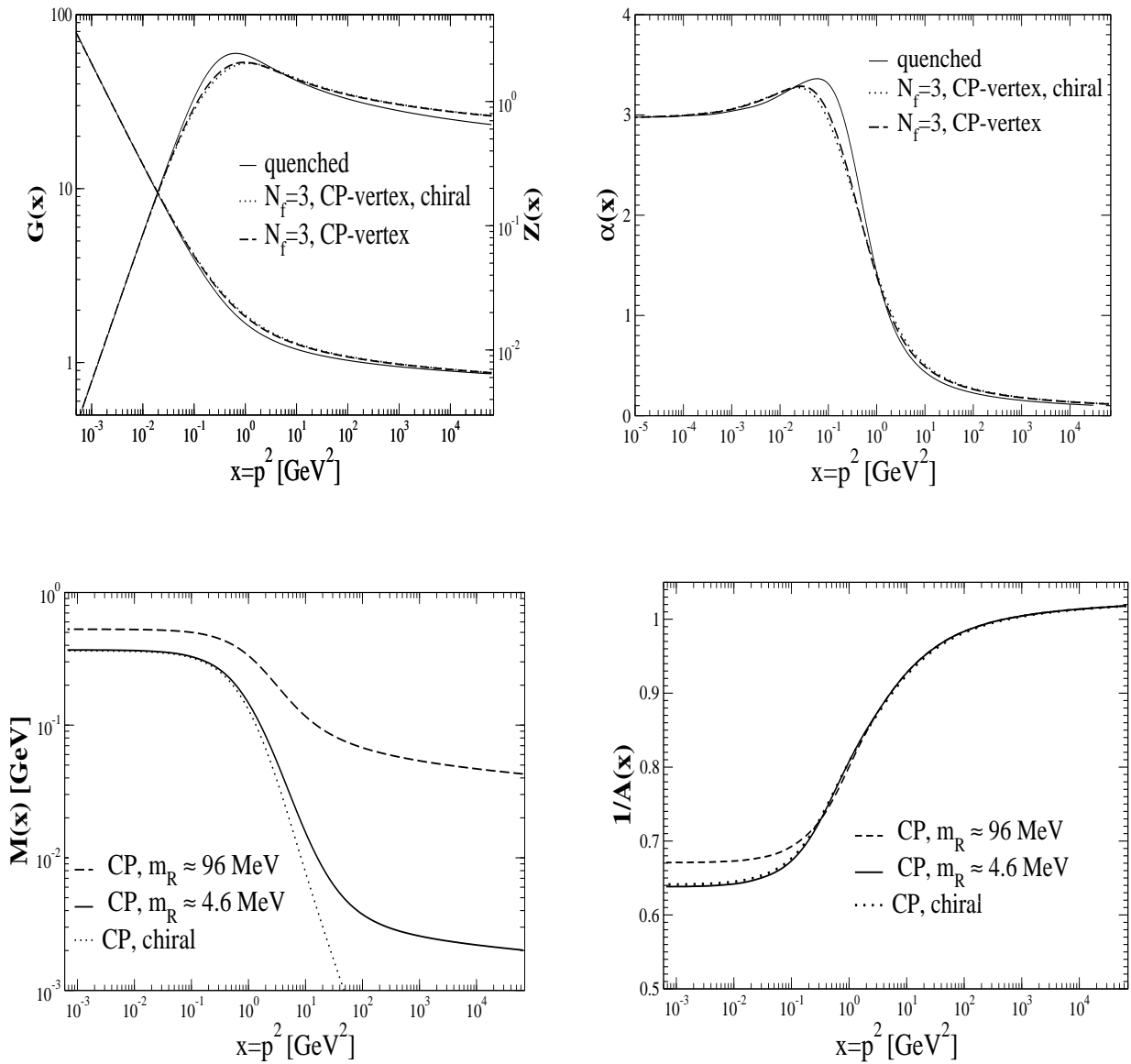

\vspace{0.4cm}
\centerline{
\epsfig{file=zg.2.eps,width=7.5cm,height=7cm}
\hspace{0.5cm}
\epsfig{file=alpha.2.eps,width=7.5cm,height=7cm}
}
\vspace{1.3cm}
\centerline{
\epsfig{file=M.2.eps,width=7.5cm,height=7cm}
\hspace{1cm}
\epsfig{file=1A.2.eps,width=7.5cm,height=7cm}
}
\caption{\sf \label{pic2.dat} Results from the unquenched calculation with $N_f=3$ 
massive quarks. We used the renormalized masses $m_{u/d}(2\,\mbox{GeV}) \approx 4.6$ 
MeV and $m_{s}(2\,\mbox{GeV}) \approx 96$ MeV. Again we chose $d=0$ for both vertices.
}
\end{figure}
%%%%%%%%%%%%%%%%%%%%%%%%%%%%%%%%%%%%%%%%%%%%%%%%%%%%%%%%%%%%%%%%%%%%%%%%%%%%%%%%%%%%

In table \ref{unqtable} we compare results for the quenched and unquenched system of 
equations. The quark mass, the pion decay constant and the chiral condensate differ 
only slightly for each vertex 
construction respectively. The only sizeable difference occurs in the running coupling.
As expected from perturbation theory the unquenched running for $N_f = 3$ results in
larger values of the running coupling at $p^2=(M_Z)^2$ compared to the quenched case 
$N_f=0$. We obtain $\alpha(M_z) \approx 0.140$, which is somewhat larger than usually 
quoted values from experiment. However, such large values are not yet excluded 
by experiment. A recent analysis of experimental data from $\tau$-decay suggests 
$\alpha(M_z) \approx 0.129$ \cite{Geshkenbein:2002ri}. 
If we increase the number of flavors in our calculation we encounter large 
numerical uncertainties and do not obtain convergence for $N_f \ge 5$. This might be 
a signal that in the range $3 < N_f \le 5$ the above discussed transition of the system 
to the chirally symmetric phase takes place. 
  
All employed vertex constructions allow for nontrivial solutions of the quark equation 
corresponding to dynamical chiral symmetry breaking. However, similar to the quenched 
case and in accordance with the results of refs.~\cite{Smekal,Ahlig}
the bare vertex construction and the CP-type vertex with $d=\delta$ generate 
much too small quark masses compared with typical phenomenological values.
For $d=0$ we obtain good results for the quark mass, the pion decay 
constant and the chiral condensate, whereas the choice $d=0.1$ leads to somewhat large 
values. It is interesting to note that $d=0$ of all values is preferred as in this case 
the quark equation resembles most the fermion equation of QED. 

In Fig.~\ref{pic1.dat} we display the ghost, gluon and quark dressing functions 
corresponding to the unquenched and quenched cases with $d=0$ from table 
\ref{unqtable} \footnote{Graphs for the choices $d=\delta$ and $d=0.1$ can be
found in ref.~\cite{Fischer}.}. 
We find different anomalous dimensions in the ultraviolet corresponding 
to the change from $N_f=0$ to $N_f=3$, {\it c.f.} eqs.~(\ref{Z_UV}), (\ref{chiral-M_UV}), 
(\ref{anom_dim}). As expected from the infrared analysis of the quark-loop the 
back-reaction of dynamical quarks in the gluon equation does not affect the infrared 
behavior of the ghost and gluon dressing functions. Consequently the infrared fixed 
point of the running coupling is the same as in pure Yang-Mills theory. 

Our results for the case of explicitly broken chiral symmetry are shown in 
Fig.~\ref{pic2.dat}. We choose $N_f=3$ with renormalized quark masses corresponding to 
$m_{u/d}(2 \,\mbox{GeV})=4.6$ MeV and $m_{s}(2 \,\mbox{GeV})=96$ MeV
within our renormalization scheme.
These masses are well in the range suggested by the Particle Data Group 
\cite{Hagiwara:2002pw}, however, they should not be identified directly as
the PDG employs an $\overline{MS}$-scheme.
Compared to the chiral case the behavior of the ghost and 
gluon dressing functions hardly changes. For the quark mass function we obtain the 
irregular asymptotic solution in the ultraviolet as expected.  

For further use, {\it e.g.} in phenomenological calculations,
we provide fits to our results for the quark propagator employing the
fit functions
\setlength{\jot}{3mm}
\beqa
M(x) &=& \frac{1}{g_1+(x/\Lambda^2_{QCD})^{g_2}} \Bigg(g_1 \, M(0) + \nonumber\\
&& \left.\hspace*{2cm}\hat{m} \left[\frac{2}{\ln(x/\Lambda^2_{QCD})}-
\frac{2}{(x/\Lambda^2_{QCD})-1}\right]^{\gamma_m} 
(x/\Lambda^2_{QCD})^{g_2} \right)\,, \hspace*{1cm}
\\
\left[A(x,s)\right]^{-1} &=& \frac{\left[A(0,s)\right]^{-1} + h_1\,(x/\Lambda^2_{QCD}) 
+ h_2 \,(x/\Lambda^2_{QCD})^2}{1 + h_3\,(x/\Lambda^2_{QCD}) + h_4 \,
(x/\Lambda^2_{QCD})^2}\,, 
\eeqa
with $x=p^2$ and the six parameters $g_1,g_2,h_1,h_2,h_3,h_4$. We used the 
renormalization point independent current-quark mass $\hat{m}$, which is related to the 
renormalized mass $M(s)$ by
\beq
\hat{m}=M(s)\, \left(\frac{1}{2}\ln[s/\Lambda^2_{QCD}]\right)^{\gamma_m}\,,
\eeq
to one loop order. For the running coupling, the ghost and the gluon dressing function 
we use the form 'Fit B', given in eq.~(\ref{fitB}) and the fit functions from 
eqs.~(\ref{zg-fit}). In table \ref{fit-table} we give our values for all parameters 
as well as the numerical results for $M(0)$ and $[A(0)]^{-1}$. Note that the scale 
$\Lambda_{QCD}^{MOM}$ is different to the corresponding scale in the chiral limit due 
to the different ultraviolet behavior of the quark-loop when quarks with non-vanishing 
bare masses are employed. When plotted the fits are virtually indistinguishable from 
our results in Fig.~\ref{pic2.dat}.
\setlength{\jot}{0mm}

%%%%%%%%%%%%%%%%%%%%%%%%%%%%%%%%%%%%%%%%%%%%%%%%%%%%%%%%%%%%%%%%%%%%%%%%%%%%%%%%%%%%%%%
\begin{table}[t]
\begin{tabular}{c||c|c|c|c|c||c|c|c|c|c||}
&\hspace*{3mm}$\Lambda^{MOM}_{QCD}$\hspace*{3mm}{ }
&\hspace*{3mm}a\hspace*{3mm}{ }&\hspace*{3mm}b\hspace*{3mm}{ }
&\hspace*{3mm}c\hspace*{3mm}{ }&\hspace*{3mm}d\hspace*{3mm}{ }
&\hspace*{3mm}$m_R$\hspace*{3mm}{ }&\hspace*{3mm}$\hat{m}$\hspace*{3mm}{ }
&\hspace*{3mm}M(0)\hspace*{3mm}{ }
&\hspace*{3mm}$g_1$\hspace*{3mm}{ }&\hspace*{3mm}$g_2$\hspace*{3mm}{ } \\
&[MeV]                & & & & &[MeV]&[MeV]    &[MeV]& &  \\\hline
\mbox{}  &625&1.22&1.00&1.33&2.01&4.6&4.6&369&1.83&1.23\\
               &   &    &    &    &    &96 &98&528&2.96&1.03\\
\end{tabular}

\vspace*{1cm}
\begin{tabular}{c||c|c|c|c|c||}
&\hspace*{3mm}$A^{-1}(0,s)$\hspace*{3mm}{ }
&\hspace*{5mm}$h_1$\hspace*{5mm}{ }&\hspace*{5mm}$h_2$\hspace*{5mm}{ }
&\hspace*{5mm}$h_3$\hspace*{5mm}{ }&\hspace*{5mm}$h_4$\hspace*{5mm}{ }\\
&    &     &&     &     \\\hline
\mbox{}  &0.638&0.515&0.00688&0.562&0.00681\\
         &0.671&0.302&0.00139&0.318&0.00137\\
\end{tabular}
\caption{\sf \label{fit-table} Parameters for the fits to the unquenched results with 
$N_f=3$, $\delta=-0.25$, $\gamma_m=12/27$ and $\beta_0=27/3$, using
the CP-vertex with $d=0$. The renormalization point $s=497 \mbox{GeV}^2$ is determined
by the condition $\alpha(s)=0.2$.
Note the change in $\Lambda_{QCD}^{MOM}$ as compared to the case of
chiral quarks. }
\end{table}
%%%%%%%%%%%%%%%%%%%%%%%%%%%%%%%%%%%%%%%%%%%%%%%%%%%%%%%%%%%%%%%%%%%%%%%%%%%%%%%%%%%%%%
\begin{figure}[th]
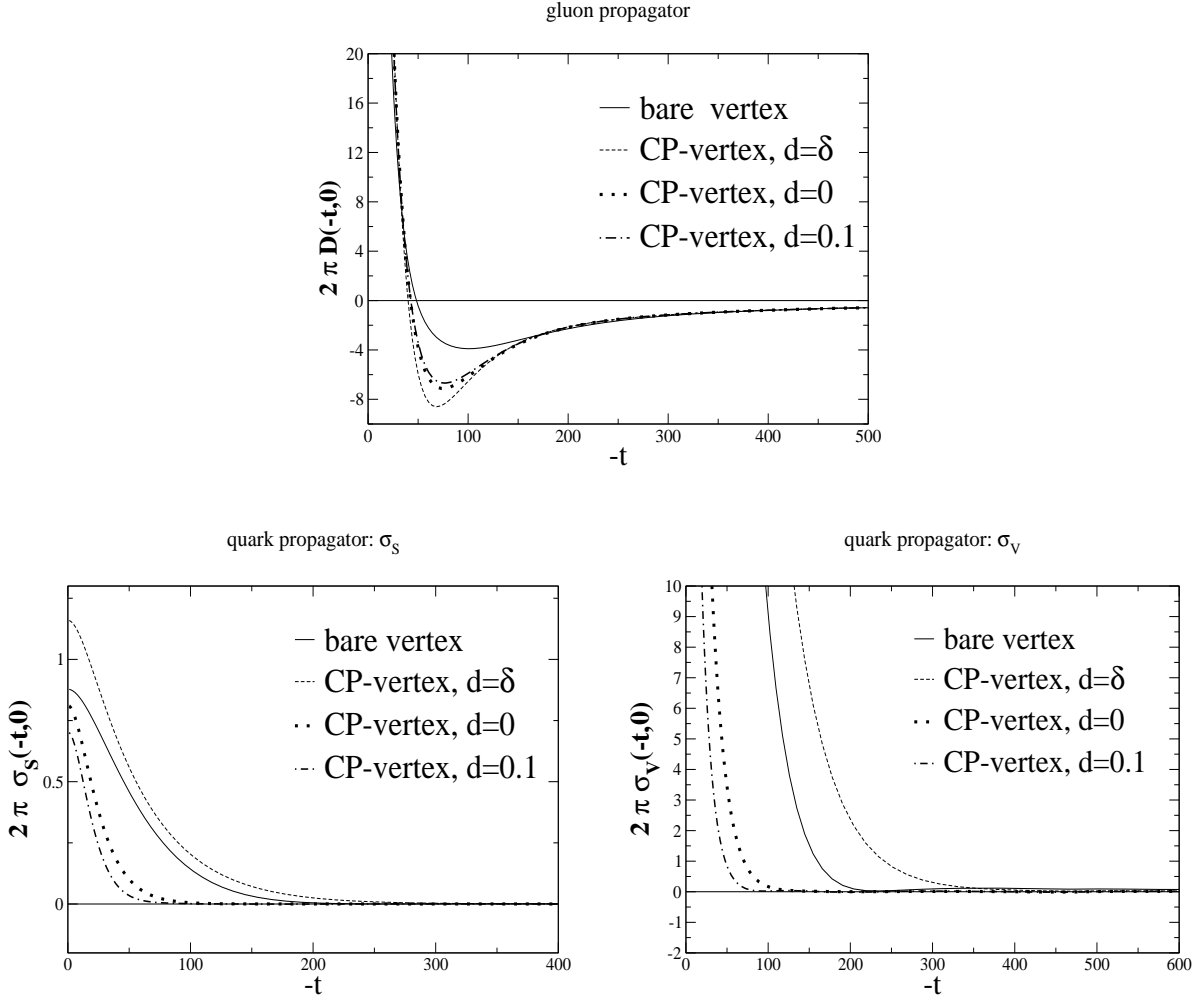

\vspace{0cm}
\centerline{
\epsfig{file=pos.glue.eps,width=7.5cm,height=6.2cm}
%\hspace{0.5cm}
%\epsfig{file=pos.peter.eps,width=7.5cm,height=6.2cm}
}
\vspace{0.8cm}
\centerline{
\epsfig{file=pos.quarkB.eps,width=7.5cm,height=6.2cm}
\hspace{0.5cm}
\epsfig{file=pos.quarkA.eps,width=7.5cm,height=6.2cm}
}
\caption{\sf \label{pic3.dat} Here we display the one dimensional Fourier transforms of 
the gluon propagator, $D(-t,\vec{p}^2)$, and the scalar and vector parts of the quark 
propagator, $\sigma_S(-t,\vec{p}^2)$ and $\sigma_V(-t,\vec{p}^2)$. 
For the three momentum we chose $\vec{p}^2=0$. The time $t$ is given in internal units.
We observe violation of reflection positivity for the gluon propagator  
but not for the quark propagator.}
\end{figure}
%%%%%%%%%%%%%%%%%%%%%%%%%%%%%%%%%%%%%%%%%%%%%%%%%%%%%%%%%%%%%%%%%%%%%%%%%%%%%%%%%%%%%%%%

Unquenched lattice calculations employing dynamical quarks are complex and time consuming
\cite{Aoki:2000kp}. To our knowledge such simulations for the propagators of QCD have 
not yet been performed. From our results in the Dyson-Schwinger approach we do not 
expect drastic differences between quenched and unquenched propagators on the lattice. 
Note, however, that our calculation includes quark-loop corrections to the gluon
self energy but not higher order vertex corrections like mesonic loops. In a model
calculation of the pion charge radius such loops have been estimated to contribute
roughly at the order of ten percent \cite{Alkofer:gu}.

Finally, we investigate possible positivity violations in the gluon and quark propagators.
According to the Osterwalder-Schrader axiom of reflection positivity a two-point correlation 
function $S$ of Euclidean field theory has to fulfill the condition \cite{Haag:1992hx}
\beq
\int_0^\infty dt \; dt^\prime \; \bar{f}(t^\prime,\vec{p}) \; S(-(t+t^\prime),\vec{p}) \; 
f(t,\vec{p}) \; > 0 \label{pos-res}
\eeq
if a physical particle is described. Here $f$ are complex valued test functions. A violation 
of this condition signals the absence of the corresponding particle from the physical 
spectrum of the theory, {\it i.e.} the particle is confined. The one-dimensional Fourier 
transform $S(t,\vec{p})$ of the propagator $S(p_0,\vec{p})$ is given by
\beq
S(t,\vec{p}) := \int \frac{dp_0}{2 \pi} \,S(p_0,\vec{p}) \,e^{i p_0 t}.
\eeq
Provided there is a region around $t_0$ where $S(-t_0,\vec{p}) <0$ one can choose a real 
test function $f(t)$ which peaks strongly at $t_0$ to show positivity violation. In the 
following we chose $\vec{p}=0$. 

In the first diagram of Fig.~\ref{pic3.dat} we display the Fourier transform of the nontrivial 
part $D(p^2)=Z(p^2)/p^2$ of the gluon propagator. Clearly one observes negative values on a 
large interval. The resulting positivity violation for the transverse gluon propagator in 
Landau gauge is a clear signal for gluon confinement. This corroborates previous findings
in the quenched approximation \cite{vonSmekal:1997is}. These positivity violations have also
been observed in lattice studies, see ref.~\cite{Langfeld:2001cz} for recent corresponding results
or ref.~\cite{Mandula:nj} for a review.

In the quark propagator positivity violations have been found in model studies, which 
solve the quark Dyson-Schwinger equation with an {\it ansatz} for the gluon propagator as 
input (see \cite{Bender:1996bm,Bender:1997jf,Roberts:2000aa} and references therein). 
Similar violations have been found in (2+1)-dimensional QED \cite{Maris:1995ns}. 
For the model of ref.~\cite{Alkofer:2002bp} or a propagator with complex conjugate poles
\cite{Stingl:1996nk} we tested our numerical routines and found positivity violations
very easily. Contrary to these findings we do not observe positivity violations 
for the quark propagator from the coupled set of DSEs. The lower panel of Fig.~\ref{pic3.dat} 
shows our results for the Fourier transform of the vector part 
$\sigma_V(p^2) = A(p^2)/(p^2 A(p^2) + B(p^2))$ and the scalar 
part $\sigma_S(p^2) = A(p^2)/(p^2 A(p^2) + B(p^2))$ of our solutions for the quark propagator 
employing four different quark-gluon vertices. All our solutions appear to be positive
definite at the present level of numerical accuray. However, a more accurate study is 
required to settle this point \cite{prep}. 
Furthermore note that even if confirmed  
our findings are not in contradiction with the absence of quarks from the physical spectrum of 
QCD as violation of positivity is a sufficient but not a necessary condition for confinement.

\section{Summary}

We have presented solutions of the (truncated) Dyson-Schwinger equations for the propagators 
of Landau gauge QCD. We first concentrated on the Dyson-Schwinger equation for the quark 
propagator. We proposed several {\it ans\"atze} for the quark-gluon vertex which consist of an 
Abelian part carrying the tensor structure of the vertex and a non-Abelian multiplicative 
correction. Our guiding principles for the construction of these vertices have been two important 
conditions on the truncated quark equation: it should be multiplicatively renormalizable and 
recover perturbation theory for large external momenta. In our truncation scheme the quark mass 
function is, as required from general arguments, 
independent of the renormalization point and has the correct asymptotic behavior 
for large momenta. 

In the quark equation both the ghost and gluon dressing function show up at least implicitly. 
In quenched approximation, which is suitable to compare to lattice results, we employ solutions 
of the ghost and gluon Dyson-Schwinger equations taken from ref.~\cite{Fischer:2002hn}. In
a second step we included the back-reaction of the quarks on the ghost and gluon system and solved
the quark, gluon and ghost Dyson-Schwinger equations self-consistently.

All our solutions exhibit dynamical chiral symmetry breaking. However, only carefully constructed 
vertex {\it ans\"atze} have been able to generate masses in the typical phenomenological range 
of $300-400$ MeV. Constructions with an Abelian part satisfying the Abelian Ward-identity are 
superior to other vertex {\it ans\"atze}. We obtained very good results for the quark mass, the 
pion decay constant and the chiral condensate by employing a generalized Curtis-Pennington 
\cite{Curtis:1990zs} vertex. In the chiral limit both, the quark mass function and the vector 
self energy coincide with recently 
obtained lattice results \cite{Bonnet:2002ih,Bowman:2002kn} 
within the numerical uncertainty. 
This agreement confirms the quality of our truncation and in turn shows 
that chiral extrapolation on the lattice works well. 

In the unquenched case including the quark-loop in the gluon equation with $N_f=3$ light quarks 
we obtain only small corrections compared to the quenched calculations. In particular for the case
of dynamically generated quark masses the quark-loop 
turns out to be suppressed in the gluon equation for small momenta. We thus showed on the level 
of our truncation that the Kugo-Ojima confinement criterion 
\cite{Kugo:1979gm, Kugo:1995km} and Zwanziger's horizon condition 
\cite{Zwanziger:2001kw, Zwanziger:1992ac} are 
satisfied in Landau gauge QCD. 

Furthermore we searched for positivity violations in the gluon and quark propagators. We confirmed 
previous findings \cite{vonSmekal:1997is} that the gluon propagator shows violation of 
reflection positivity. Thus the gluon
is not contained in the physical state space of QCD. We did not find similar violations for the quark 
propagator. This issue is currently investigated in a more detailed study \cite{prep}.

\section*{Acknowledgments}
We are indebted to P.~Maris and P.~Watson for a critical reading of the manuscript and 
useful comments. We are grateful to P.~Bowman for communicating lattice data.
We thank S.~Ahlig, J.~Bloch, K.~Langfeld, H.~Reinhardt, 
C.~Roberts, L.~von Smekal, P.~Tandy and A.~Williams
for helpful discussions.\\
This work has been supported by the DAAD and the DFG under contracts Al 279/3-4 and 
GRK683 (European graduate school T\"ubingen--Basel).

\begin{appendix}
\section{Appendix A: Angular integrals}
The angular integrals employed in the ultraviolet analysis are given by
\beqa
\int_0^\pi d\theta \frac{\sin^2(\theta)}{z^2} &=& \frac{\pi}{2} 
       \left[\frac{\Theta(x-y)}{x(x-y)} + \frac{\Theta(y-x)}{y(y-x)} \right]  \\
\int_0^\pi d\theta \frac{\sin^2(\theta)}{z} &=& \frac{\pi}{2} 
       \left[\frac{\Theta(x-y)}{x} + \frac{\Theta(y-x)}{y} \right]  \\
\int_0^\pi d\theta \sin^2(\theta) &=&  \frac{\pi}{2} \\
\int_0^\pi d\theta \sin^2(\theta) \: z &=&  \frac{\pi}{2} (x+y) \\
\int_0^\pi d\theta \sin^2(\theta) \: z^2 &=&  \frac{\pi}{2} \left((x+y)^2 + xy \right)
\eeqa
where the squared momentum $z$ is defined as $z=(p-q)^2 = x+y-2\sqrt{xy} \cos(\theta)$.     
\end{appendix}

\end{document}